%

\input harvmac.tex
\overfullrule=0mm
\hfuzz 15pt

\input epsf.tex
\newcount\figno
\figno=0
\def\fig#1#2#3{
\par\begingroup\parindent=0pt\leftskip=1cm\rightskip=1cm\parindent=0pt
\baselineskip=11pt
\global\advance\figno by 1
\midinsert
\epsfxsize=#3
\centerline{\epsfbox{#2}}
\vskip 12pt
{\bf Fig. \the\figno:} #1\par
\endinsert\endgroup\par
}
\def\figlabel#1{\xdef#1{\the\figno}}
\def\encadremath#1{\vbox{\hrule\hbox{\vrule\kern8pt\vbox{\kern8pt
\hbox{$\displaystyle #1$}\kern8pt}
\kern8pt\vrule}\hrule}}


%
%
\def\za{\alpha} \def\zb{\beta} \def\zg{\gamma} \def\zd{\delta}
   
\def\zl{\lambda} \def\zm{\mu} \def\zn{\nu} \def\zo{\omega}
 \def\zr{\rho} \def\zs{\sigma}

\def\bl{\bar{\lambda}}

\def\zG{\Gamma}   \def\zF{\Phi}
\def\zL{\Lambda} \def\zO{\Omega} \def\zS{\Sigma}

\def\IZ{Z\!\!\!Z}
\def\dC{C\kern-6.5pt I}

\def\la{\langle} \def\ra{\rangle}

\def\Exp{{\rm Exp}}

  \def\zF{\Phi} \def\zL{\Lambda}
\def\zO{\Omega}\def\zG{\Gamma}

\def\De{D_{{\rm even}}}

\def\la{\langle} \def\ra{\rangle}

\def\ba{\bar a} \def\bb{\bar b}

\def\IZ{Z\!\!\!Z}
\def\cB{{\cal B}}
\def\cU{{\cal U}}


%
%
\def\frac#1#2{{\scriptstyle{#1 \over #2}}}                
\def\inv#1{\scriptstyle{1 \over #1}}

%
%
\def\CA{{\cal A}}       \def\CB{{\cal B}}       
\def\CD{{\cal D}}       \def\CE{{\cal E}}       
\def\CG{{\cal G}}              
              
       \def\CN{{\cal N}}       
\def\CP{{\cal P}}              
              
\def\CV{{\cal V}}              

\def\({ \left( }\def\[{ \left[ }
\def\){ \right) }\def\]{ \right] }
%


\def\IR{\relax{\rm I\kern-.18em R}}
\font\cmss=cmss10 \font\cmsss=cmss10 at 7pt
\def\IZ{\relax\ifmmode\mathchoice
{\hbox{\cmss Z\kern-.4em Z}}{\hbox{\cmss Z\kern-.4em Z}}
{\lower.9pt\hbox{\cmsss Z\kern-.4em Z}}
{\lower1.2pt\hbox{\cmsss Z\kern-.4em Z}}\else{\cmss Z\kern-.4em Z}\fi}
\def\inbar{\,\vrule height1.5ex width.4pt depth0pt}
\def\IB{\relax{\rm I\kern-.18em B}}
\def\ID{\relax{\rm I\kern-.18em D}}
\def\IE{\relax{\rm I\kern-.18em E}}
\def\IF{\relax{\rm I\kern-.18em F}}
\def\IG{\relax\hbox{$\inbar\kern-.3em{\rm G}$}}
\def\IH{\relax{\rm I\kern-.18em H}}
\def\II{\relax{\rm I\kern-.18em I}}
\def\IK{\relax{\rm I\kern-.18em K}}
\def\IL{\relax{\rm I\kern-.18em L}}
\def\IM{\relax{\rm I\kern-.18em M}}
\def\IN{\relax{\rm I\kern-.18em N}}
\def\IO{\relax\hbox{$\inbar\kern-.3em{\rm O}$}}
\def\IP{\relax{\rm I\kern-.18em P}}
\def\IQ{\relax\hbox{$\inbar\kern-.3em{\rm Q}$}}
\def\IGa{\relax\hbox{${\rm I}\kern-.18em\Gamma$}}
\def\IPi{\relax\hbox{${\rm I}\kern-.18em\Pi$}}
\def\ITh{\relax\hbox{$\inbar\kern-.3em\Theta$}}
\def\IOm{\relax\hbox{$\inbar\kern-3.00pt\Omega$}}

\def\Z{\IZ}


\def\d{{\rm d}}

\def\bv{\bigg\vert}\def\un{{\bf 1}}

\def\Ga{\alpha}
\def\Gd{\delta}

\def\Gk{\kappa}\def\Gl{\lambda}\def\GL{\Lambda}
\def\Gm{\mu}\def\Gn{\nu}
\def\Gr{\rho}
\def\Gs{\sigma}\def\Gt{\tau}


\def\mod{{\rm mod\,}}

\def\nind{\noindent}

\def\rep{representation}

\def\slh{\widehat{sl}}\def\soh{\widehat{so}}
\def\av{a^{\vee}}
\def\bv{b^{\vee}}

\def\\#1 {{\tt\char'134#1} }
\def\M{{\widetilde M}}
\def\a{t}\def\b{r}\def\bb{\bar\b}\def\ba{\bar\a}
\def\c{s}\def\d{u}

\catcode`\@=11
\def\Eqalign#1{\null\,\vcenter{\openup\jot\m@th\ialign{
\strut\hfil$\displaystyle{##}$&$\displaystyle{{}##}$\hfil
&&\qquad\strut\hfil$\displaystyle{##}$&$\displaystyle{{}##}$
\hfil\crcr#1\crcr}}\,}   \catcode`\@=12

\def\encadre#1{\vbox{\hrule\hbox{\vrule\kern8pt\vbox{\kern8pt#1\kern8pt}
\kern8pt\vrule}\hrule}}
\def\encadremath#1{\vbox{\hrule\hbox{\vrule\kern8pt\vbox{\kern8pt
\hbox{$\displaystyle #1$}\kern8pt}
\kern8pt\vrule}\hrule}}

\newdimen\xraise\newcount\nraise
\def\xpoint{\hbox{\vrule height .45pt wNth .45pt}}
\def\udiag#1{\vcenter{\hbox{\hskip.05pt\nraise=0\xraise=0pt
\loop\ifnum\nraise<#1\hskip-.05pt\raise\xraise\xpoint
\advance\nraise by 1\advance\xraise by .4pt\repeat}}}
\def\ddiag#1{\vcenter{\hbox{\hskip.05pt\nraise=0\xraise=0pt
\loop\ifnum\nraise<#1\hskip-.05pt\raise\xraise\xpoint
\advance\nraise by 1\advance\xraise by -.4pt\repeat}}}
\def\bdiamond#1#2#3#4{\raise1pt\hbox{$\scriptstyle#2$}
\,\vcenter{\vbox{\baselineskip12pt
\lineskip1pt\lineskiplimit0pt\hbox{\hskip10pt$\scriptstyle#3$}
\hbox{$\udiag{30}\ddiag{30}$}\vskip-1pt\hbox{$\ddiag{30}\udiag{30}$}
\hbox{\hskip10pt$\scriptstyle#1$}}}\,\raise1pt\hbox{$\scriptstyle#4$}}



\def\IC{\relax\hbox{$\inbar\kern-.3em{\rm C}$}}

\def\msy{y }
\ifx\msan\msy
\input amssym.def
\input amssym.tex
\def\IZ{\Bbb Z}\def\IR{\Bbb R}\def\IC{\Bbb C}\def\IN{\Bbb N}
\def\gg{\goth g}
\else
\def\gg{g}
\fi
\def\ggh{\hat\gg}




\lref\FL{V.A. Fateev and S.L. Lukyanov, {\it Int. J. Mod. Phys.}
{\bf A3} (1988) 507.}


\lref\BBSW{F.A. Bais and P.G. Bouwknegt, {\it Nucl. Phys.} {\bf B279}
(1987) 561 \semi
A.N. Schellekens and N.P. Warner, {\it Phys. Rev. D} {\bf
34} (1986) 3092.}

\lref\KS{V.G. Kac and M.N. Sanielevici, {\it Phys. Rev. D} {\bf
37} (1988) 2231.}
\lref\KW{V.G. Kac and M. Wakimoto, {\it Adv. in Math.} {\bf 70}
(1988) 156.}


\lref\B{D. Bernard, {\it  Nucl. Phys.} {\bf B288} (1987) 628.}
\lref\ChR{P. Christe and F. Ravanini, {\it Int. J. Mod. Phys.}
{\bf A4} (1989) 897. }

\lref\GRW{T. Gannon, P. Ruelle and M.A. Walton, Automorphism
Modular Invariants of Current Algebras, preprint LETH-PHY-3/95,
US-FT/22-94, UCL-IPT-95/4, hep-th/9503141.}

\lref\CIZ{A. Cappelli, C. Itzykson and J.-B. Zuber,
{\it Nucl. Phys.} {\bf B280} [FS18] (1987)
445; {\it Comm. Math. Phys.} {\bf 113} (1987) 1
\semi A. Kato, {\it Mod. Phys. Lett.} {\bf A2} (1987) 585.}

\lref\IDG{C. Itzykson, Nucl. Phys. (Proc. Suppl.) {\bf 5B} (1988) 150
\semi P. Degiovanni, Comm. Math. Phys. {\bf 127} (1990) 71. }

\lref\Gan{T. Gannon, {\it Comm. Math. Phys.} {\bf 161} (1994) 233;
{\it The Classification of $SU(3)$ Modular Invariants Revisited},
hep-th 9404185.}

\lref\Gancos{T. Gannon and M.A. Walton,
{\it  On the Classification of
Diagonal Coset Modular Invariants}, hep-th 9407055.  }

\lref\ScY{A.N. Schellekens and S. Yankielowicz,
 {\it Phys. Lett.} {\bf  B227} (1989) 387.}
\lref\SY{A.N. Schellekens and S. Yankielowicz, {\it Nucl. Phys.}
{\bf B334} (1990) 67.}
\lref\SYa{A.N. Schellekens and S. Yankielowicz, {\it Nucl. Phys.}
{\bf B327} (1989) 673.}

\lref\EV{E. Verlinde, {\it  Nucl. Phys.} {\bf B300} [FS22] (1988) 360. }
\lref\MSDV{R. Dijkgraaf and E. Verlinde, {\it Nucl. Phys.} (Proc. Suppl.)
{\bf 5B}  (1988) 87.}

\lref\MS{G. Moore and N. Seiberg,  {\it Nucl. Phys.} {\bf B313}
(1989) 16; {\it Comm. Math. Phys.} {\bf 123} (1989) 177.}


\lref\BYZ{R. Brustein, S. Yankielowicz and J.-B. Zuber, {\it Nucl.
Phys.} {\bf B313}  (1989) 321.}

\lref\PZ{V.B. Petkova and J.-B. Zuber, {\it Nucl. Phys.} {\bf B438}
(1995) 347. }


\lref\VPun{V. Pasquier, {\it Nucl. Phys.} {\bf B285} [FS19] (1987) 162
\semi V. Pasquier, {\it J. Phys.} {\bf A20} (1987) 5707.}
\lref\Ko{I.K. Kostov, {\it Nucl. Phys.} {\bf B 300} [FS22] (1988) 559.}

\lref\DFZun{P. Di Francesco and J.-B. Zuber,
{\it Nucl. Phys.} {\bf B338} (1990) 602
.}
\lref\DFZ{P. Di Francesco and J.-B. Zuber,
in {\it Recent Developments in Conformal Field Theories}, Trieste Conference
1989, S. Randjbar-Daemi, E. Sezgin and J.-B. Zuber eds., World Scientific
1990 \semi
P. Di Francesco, {\it Int. J. Mod. Phys.} {\bf A7} (1992) 407.}

\lref\GHJ{F.M. Goodman, P. de la Harpe and V.F.R. Jones, {\it Coxeter
Graphs and Towers of Algebras}, Springer-Verlag, Berlin (1989).}

\lref\RST{K.-H. Rehren, Ya.S. Stanev and I.T. Todorov,
{\it Characterizing invariants for local extensions of current algebras},
preprint DESY 94-164 ESI 132, hep-th/ 9409509, {\it Comm. Math. Phys.} (1995)
to appear. }

\lref\BI{E. Bannai, T. Ito, {\it Algebraic Combinatorics I: Association
Schemes}, Benjamin/Cummings (1984).}

\lref\Ocn{A. Ocneanu, communication at the Workshop
{\it Low Dimensional Topology, Statistical Mechanics and Quantum Field Theory},
 Fields Institute, Waterloo, Ontario, April 26--30, 1995.}

\lref\So{N. Sochen, {\it Nucl. Phys.} {\bf B360} (1991) 613. }

\lref\Zub{J.-B. Zuber, {\it Graphs and Reflection Groups}, SPhT 95/089,
hep-th 9507057,  {\it Comm. Math. Phys.} to appear. }



\Title{
ASI-TPA/14/95
\quad SPhT 95/118
\qquad{\tt hep-th/9510175  }}
{{\vbox {
\vskip-10mm
\centerline{From CFT to Graphs }
}}}
\medskip
\centerline{V.B. Petkova}
\bigskip
\centerline{\it Arnold Sommerfeld Institute for Mathematical Physics,}
\centerline{\it
TU Clausthal, Leibnizstr. 10, D-38678
Clausthal--Zellerfeld, Germany,}
\medskip
\centerline{  \it  Institute for Nuclear Research
and Nuclear Energy
\footnote{${}^\dagger$}{Permanent address.},}
\centerline{\it Tzarigradsko Chaussee 72, 1784 Sofia, Bulgaria
}
\bigskip
\centerline{and}
\medskip
\centerline{J.-B. Zuber}\bigskip

\centerline{ \it CEA, Service de Physique Th\'eorique de Saclay
\footnote*{Laboratoire de la Direction des Sciences
de la Mati\`ere du Commissariat \`a l'Energie Atomique.},}
\centerline{ \it F-91191 Gif sur Yvette Cedex, France}

\vskip .2in

\noindent 
In this paper, we pursue the discussion of the connections
between rational conformal field theories (CFT) and graphs. We
generalize our recent work on the relations of operator product
algebra (OPA) structure constants of $sl(2)\,$ theories  with the
Pasquier algebra attached to the graph. We show that in a variety
of CFT built on $sl(n)\,$ -- typically  conformal embeddings and
orbifolds, similar considerations enable one to write a linear
system satisfied by the matrix elements of the Pasquier algebra
in terms of conformal data -- quantum dimensions and fusion
coefficients.  In some cases, this provides a sufficient
information for the determination of all the eigenvectors of an
adjacency matrix, and hence of a graph.
\bigskip

\Date{10/95\qquad\qquad submitted to Nuclear Physics B}
%


\newsec{Introduction}
\nind In this paper we explore further the connections between
rational conformal field theories and graphs.
The idea inherited from the earlier work of Pasquier \VPun\ is
that there are tight connections between some features of
rational conformal field theories based on $sl(n)\,,$ typically
WZW models and the  corresponding minimal cosets, on the one hand,
and generalized lattice ``height'' models based on graphs,
on the other. The most conspicuous of these
connections lies in the {\it spectral} properties of the two
kinds of theories: the same $sl(n)$ weights that label the spin
zero fields of the CFT also label the eigenvalues of the
adjacency matrices of the graphs \DFZun.

This connection has been strengthened  by the recent
observation \PZ\ that it extends to the level of operator algebras:
in $sl(2)$ theories, the
ratios of structure constants of the OPA of spin zero fields
pertaining to a pair of CFT's of identical level are given by
the structure constants $M_{\Gl\Gm}^{\ \ \Gn}$
of an algebra intrinsically attached to the graph.
This algebra, that we call the Pasquier algebra, has structure
constants given by a Verlinde-like formula, in which the
components of the eigenvectors $\psi$ of the adjacency matrix of the
graph replace the modular $S$ matrix: see below for a more
precise expression.

These considerations assumed the knowledge of the graph. In \DFZun,
graphs relevant for the case of $sl(3)$ were found
by empiric ways, essentially by trials and errors. The purpose of
this paper is to show that, at least in a restricted class
of CFT  with an extended chiral symmetry algebra,
the previous connection
between the structure constants and the Pasquier algebra
may be inverted. Data of the CFT
enable one to determine the structure constants $M$ of the
Pasquier algebra, the diagonalization of which yields the
eigenvectors $\psi$ and thus the adjacency matrix and the graph.

This paper is organized as follows: after introducing
in sect. 2 some notations  and some basic properties of the
graphs, we shall study in sect. 3
the operator algebra of conformal theories   associated with
conformal embeddings
and propose an equation that connects
conformal and graphical data. This will be probed in sect. 4 on
various examples. Section 5 uses the theory
of ``$C$-algebras" to obtain more explicit relations
on  eigenvectors of the adjacency matrices. Two appendices present
 one further example and a discussion of the $sl(3)$
 block-diagonal $D$ -- series.

%


\newsec{Notations}
\nind
This section is devoted to a summary of notations and concepts
introduced in our former work.

\subsec{On CFT's}
\nind Working with the $sl(n)$ algebra and its affine extension
$\slh(n)_k\,,$  at a certain integer level $k$, we first
introduce some Lie algebraic objects.
\def\GLh{\hat{\GL}}
Let $\GLh_1,\cdots,\GLh_{n-1}$ be the fundamental weights of $sl(n)$.
(The hat is intended to distinguish them from the later notation
$\GL$ for a pair of weights.)
Let $\rho=\GLh_1+\cdots +\GLh_{n-1}$ be the sum of these fundamental weights.
Since in a WZW theory, the weight $\Gr$ labels the identity field,
we shall also occasionally denote it by $\un$.
The set of integrable weights
(shifted by $\rho$) of the affine algebra $\slh(n)_k$ is
\eqn\Ia{\CP_{++}^{(k+n)}=\{\Gl=\Gl_1\GLh_1+\cdots
+\Gl_{n-1}\GLh_{n-1}|\ \Gl_i \in
\IZ\,, \quad \Gl_i\ge 1\,, \quad
\sum_i\,\Gl_i\le k+n-1\} \ .}

 We shall  encounter the $\IZ_n$  group of automorphisms
of this set of weights, generated by
\eqn\Iaa{\Gs:\ \Gl=(\Gl_1,\Gl_2,\cdots, \Gl_{n-1})\mapsto
\Gs(\Gl)= (
n+k-\Gl_1-\cdots-\Gl_{n-1}, \Gl_1,\cdots,\Gl_{n-2})\ .}
We also make use of the $n$ (linearly dependent) vectors $e_i$
\eqn\Ib{e_1=\GLh_1,\qquad e_i=\GLh_i-\GLh_{i-1},\quad i=2,\cdots
, n-1,\qquad e_n=-\GLh_{n-1}\ }
 (the weights of the fundamental \rep\ of  highest weight $\GLh_1$)
and of the symmetric bilinear form on weight space
\eqn\Iba{(e_i,e_j)= \Gd_{ij}-{1\over n}\ .}

We shall be mainly considering the WZW model with a current
algebra
$\slh{(n)}_k\,,$ or the simplest coset CFT's
$$ {\slh{(n)}_{k-1}\times \slh{(n)}_1\over \slh{(n)}_{k}}$$
also called the ``minimal $W_n$ models". In the WZW theories,
primary fields are labelled by a pair of integrable weights at level
$k$, and this pair is denoted by a capital $\GL$
\eqn\Ic{\GL=(\Gl,\bar\Gl), \qquad \Gl,\bar\Gl\in \CP_{++}^{(k+n)} \ .}
In the coset theories, we rather need two such pairs at level
$k$ and $k-1$
\foot{In principle we should use  {\it three} pairs of weights
$${\Big((\Gl,\bar\Gl),(\Gl',\bar\Gl'), (\Gl'',\bar\Gl'')\Big)
,\qquad \Gl,\bar\Gl\in \CP_{++}^{(n+k)}, \quad
 \quad \Gl',\bar\Gl'\in \CP_{++}^{(n+k-1)},\quad
\quad \Gl'',\bar\Gl''\in \CP_{++}^{(n+1)}  \,,} $$
However we can make use of the $\IZ_n$ automorphism
 $\Gs$ of \Iaa, and of the fact that in the coset theory
the pairs $(\zl,\zl')\,$ and $(\bl,\bl')\,$ are
determined modulo a diagonal action of $\Gs\,,$
   to rotate $\Gl'',\bar\Gl''$ to the standard value $\Gr$. }
$${\Big((\Gl,\bar\Gl),(\Gl',\bar\Gl')\Big)
,\qquad \Gl,\bar\Gl\in \CP_{++}^{(k+n)}, \quad
 \qquad \Gl',\bar\Gl'\in \CP_{++}^{(k+n-1)} \ .}$$

In fact, as in \PZ, we shall concentrate on the  ``thermal"
fields, for which $\Gl'=\bar\Gl'=\Gr$, and thus label simply these fields
by the same notation $\GL$ as in \Ic.

\medskip
As in  any CFT, the way the left and right
representations of the chiral algebra (affine or Virasoro or more generally
$W_n$) are coupled to produce the primary fields of the theory
is encoded in the genus one partition function
\eqn\Iea{Z= \sum_{\Gl,\bar\Gl } \CN_{\Gl\,\bar\Gl}\, \chi_{_{\scriptstyle
\Gl}}(q)\, \chi_{_{\scriptstyle \bar\Gl}}(\bar q)\ .}
Classification of such modular invariant partition functions
has been accomplished only in  a few cases,
  \refs{\CIZ \Gan{--}\IDG} .
In eq. \Iea, we have made use of notations relevant for affine
algebras.
 For coset theories, it is clear that a large class of solutions
is obtained through the factorization of the matrix $\CN$ into
those of the factors of the coset
\eqn\Ieaa{\CN^{{\rm coset}}_{(\Gl,\Gl'),(\bar\Gl,\bar\Gl')}
=\CN^{(k)}_{\Gl\bar\Gl}
\CN^{(k-1)}_{\Gl'\bar\Gl'}
}
(see  \Gancos\ for a thorough discussion of all the possible
solutions)
but we shall consider solely the cases in which the
level $(k-1)$ invariant is ``trivial"
\eqn\Ieaaa{\CN^{{\rm coset}}_{(\Gl,\Gl'),(\bar\Gl,\bar\Gl')}
=\CN^{(k)}_{\Gl\bar\Gl}
\Gd_{\Gl'\bar\Gl'} \ . }

\medskip\nind
In general,
we find it appropriate to distinguish between the two cases
of:

\item{*}{Type 
I theories} for which the partition function
\Iea\ may be recast as a sum of squares of sums of characters
\eqn\Ieb{ Z=\sum_i |\sum_{\Gl\in \CB_i} \chi_{_{\scriptstyle
\Gl}} |^2 \,. }
\item{*}{Type 
II theories} for which this rewriting is impossible
without the introduction of signs
\eqn\Iec{ Z=\sum_i \,\pm \,|\sum_{\Gl\in \CB_i}
\chi_{_{\scriptstyle \Gl}} |^2 \,.}

\nind The former case signals the existence of
an {\it extended chiral algebra}, generated by holomorphic fields
that appear in the block of the identity, and whose  irreducible
representations labelled $\CB_i$ decompose into irreducible
representations of the current algebra as shown in the expression
of $Z$.
\foot{
Some representations $\lambda$ may appear in $\CB_i\,$ with a
nontrivial multiplicity ${\rm mult}_{\CB_i}(\zl)\in \IN\,,$ i.e.,
in \Ieb\
$\chi_{_{\CB_i}}=\sum_{\zl \in \CP_{++}
}\, {\rm mult}_{\CB_i}(\zl) \,
\chi_{_{\zl}}\,.$ }
The latter cases,(that concern us less in the present paper,
are known to be related to some of the former by a twist of the
right sector with respect to the left one  according to an
automorphism of the fusion rules   \refs{\MSDV\MS{--}\GRW}.

\medskip
In many of the subsequent considerations, we shall be discussing
spin zero fields, for which the left and right components are
identical $\Gl=\bar \Gl$. For those fields, we shall adopt a
shorthand notation, using $\Gl$ instead of $\GL=(\Gl,\Gl)$.
We do of course the same for {\it chiral} quantities, i.e.
quantities intrinsically attached to a single component. This is in
particular the case of the fusion algebra whose coefficients are given by
the celebrated Verlinde formula \EV\
\eqn\Ie{ N_{\Gl\Gm}^{\ \ \Gn}=\sum_{\Gs}
{S_{\Gl\Gs}S_{\Gm\Gs}S_{\Gn\Gs}^*
\over S_{\un\Gs}} \ .}

In contrast, coefficients, or structure constants, of the
operator product algebra (OPA) are {\it not} chiral quantities
and depend in an essential way on the coupling between the left
and  right sectors.


\subsec{On Graphs}
\nind
The second category of objects that we shall be handling are
graphs. When dealing with $sl(n)$ theories, we postulate that
each of these graphs $\CG$ satisfies the following requirements:

\item{i)} it is connected;
\item{ii)} it is symmetric, i.e. unoriented;
\item{iii)} let $\CV$ be the set of vertices; to each $a\in \CV$
may be attached a $\IZ/n\IZ$ grading $\tau(a)$, the ``$n$-ality",
and the only nonvanishing entries of the adjacency matrix $G$
(i.e. the only edges)
are between vertices of different $\tau$. This enables one to
split  this adjacency matrix into a sum of $n-1$ matrices
\eqn\If{G=G_1+G_2+\cdots +G_{n-1}\,,}
where $G_p$ is the adjacency matrix describing the edges that
connect vertices of $n$-ality differing by $p$
\eqn\Ig{
 (G_p)_{ab}\ne 0 \qquad{\rm only \ if \ }\quad \tau(b)=\tau(a)+p
\ \mod n\ ;}
Accordingly, the graph may be regarded as the superposition on the
same set of vertices of $n-1$
oriented (except for $p=n/2\,$),  not   all necessarily connected,
 graphs $\CG_p$
of adjacency matrices $G_p$, $p=1,\cdots, n-1$;
\item{iv)}  there exists an involution $a\mapsto \av$ on $\CV$
such that $\tau(\av)=-\tau(a)$ and
\eqn\Iga{ (G_p)_{ab} = (G_p)_{\bv\av}\ ;}
\item{v)} the matrices $G_p$ are pairwise transposed of one another:
\eqn\Ih{{}^tG_p=G_{n-p}\,;}
\item{vi)} the matrices $G_p$ commute among themselves, in particular
each $G_p$ commutes with its transpose $G_{n-p}$, hence is ``normal",
i.e. diagonalizable in an orthonormal basis common to all of them;
\item{vii)}  these common eigenvectors are
labelled by integrable weights $\Gl\in \CP_{++}^{(k+n)}$ for some
level $k$, we  denote them $ \psi^{(\Gl)}_a$, ($a\in \CV$),
 and the corresponding eigenvalues of $G_1$, $G_2$,
\dots, $G_{n-1}$  are given by the following formulae
\eqn\Ii{
\eqalign{\gamma^{(\Gl)}_1 
&=\sum_{i=1}^n \exp -{2i\pi\over h}(e_i,\Gl) \cr
         \gamma^{(\Gl)}_2
&=\sum_{1\le i<j\le n} \exp -{2i\pi\over h}
                                        ((e_i+e_j),\Gl) \cr
                \vdots  \quad           &\qquad\quad \vdots
\qquad\qquad \vdots \cr
        \gamma^{(\Gl)}_{n-1} 
        &=\sum_{1\le i_1<\cdots i_{n-1}\le n}
        \exp -{2i\pi\over h}((e_{i_1}+\cdots + e_{i_{n-1}}),\Gl)\,,
\cr
        &\gamma^{(\Gl)}_{n-p} = \(\gamma^{(\Gl)}_{p}
        \)^* = \gamma^{(\Gl^*)}_{p} \,, \cr } }
where  $h=k+n$ and $\Gl^*\,$ is  the conjugate weight
$(\zl_{n-1}\,,..., \zl_1)\,;$  some of these $\Gl$
may occur with multiplicities larger than one;
\item{viii)}  $\Gr=(1,1,\cdots,1) \,$  is among these $\Gl\,,$
with multiplicity 1: it corresponds to the eigenvector of largest
eigenvalue, $\gamma^{(\zr)}_{p}\ge |\gamma^{(\zl)}_{p}|\,,$ the
so--called Perron--Frobenius eigenvector;
 its components  $\psi_a^{(\zr)}\,,$ $a\in \CV\,,$ are
nonvanishing and positive;
\item{ix)}   the graph $\CG_1$ admits at least one extremal
vertex, i.e. a vertex on which only one edge is ending
and from which only one edge is starting.  This vertex is
denoted
$\un\,$ and we put two additional requirements on it
\itemitem{1)} $\un^{\vee}=\un\,,$ and thus $\tau(\un)=0\,,$ \
\foot{Although assumption 1) is not satisfied by all  graphs that we know,
it seems
to be met by those that are relevant in the context of the present paper,
namely  those of type I.   }
\itemitem{2)} all the components of
the $\psi$ relative to this vertex are nonvanishing and
positive: $\forall \Gl\quad \psi^{(\Gl)}_1> 0\ .$

As a consequence of ix--1), the two points connected to 1 are
conjugate to one another under the involution and will be denoted
respectively  $a_f$ and $a_f^{\vee}$.

For the benefit of the reader who is not convinced by the
naturalness of these requirements, we shall point out that the
graphs $\CG_p$, resp. their adjacency matrices $(G_p)_{ab}$, are
natural extensions of the fusion graphs, resp. fusion
coefficients of the $n-1$ fundamental representations of
$\slh{(n)}_k$; in that particular case, $\CV=\CP_{++}^{(k+n)}$,
$\tau$  is the natural $\IZ/n\,\IZ$ grading of
representations of $sl(n)$ (number of boxes of the Young tableau
modulo $n$), while the involution is the conjugation of
representations; the commutation and spectral properties of the
$G_p$ are also natural in that case, with the expressions \Ii\
following from the Verlinde formula, together with the explicit
form of the modular matrix $S$; in that particular case,
$\psi\equiv S$.

In the case $n=2$, the only graphs satisfying these conditions
are the $ADE$ Dynkin diagrams \GHJ\ and the weights $\Gl$
labelling the eigenvalues are the Coxeter exponents. For $n=3$,
some graphs have been found satisfying these conditions, see
\DFZun. By extension of the $sl(2)$ situation, we call again
``exponents'' the corresponding $\Gl$ and denote their set by
$\Exp$.  It is a simple consequence of eq. \Ig\ that the set
$\Exp $ is invariant under the automorphism \Iaa\  and that the
eigenvectors may be chosen so as to satisfy
\eqn\quar{\psi^{(\Gs(\Gl))}_a=e^{2i\pi{\Gt(a)\over n}}
\psi^{(\Gl)}_a\ .}

These graphs are {\it not}, however, in general,  fusion graphs.
This is evidenced if out of the set of eigenvectors
$ \psi^{(\Gl)}_a\,$ and their complex conjugates  $\psi^{(\Gl)\,
*}_a\,,$ satisfying
\eqna\Ik
$$\eqalignno{
\sum_{a\in \CV} \  \psi^{(\Gl)}_a\,\psi^{(\Gn)\, *}_a
&= \Gd_{\Gl \Gn}\,, \qquad
\sum_{\Gl \in \Exp }\  \psi^{(\Gl)}_a\, \psi^{(\Gl)\, *}_c =
\Gd_{a c}\,,
&\Ik{a}\cr
  &\psi^{(\Gl)}_{a^{\vee}} =  \psi^{(\Gl)\, *}_a=
\psi^{(\Gl^*)}_a\,, & \Ik{b}\cr
}$$
one forms the following two sets of real numbers
\eqna\Ij
$$\eqalignno{ M_{\Gl\Gm}^{\ \ \Gn}&=
\sum_{a\in \CV} \  {\psi^{(\Gl)}_a\psi^{(\Gm)}_a\psi^{(\Gn)\,
*}_a \over \psi^{(\Gr)}_a\,,}
  & \Ij a  \cr
N_{ab}^{\,\, c}&=
\sum_{\Gl \in \Exp } \ {\psi^{(\Gl)}_a\psi^{(\Gl)}_b\psi^{(\Gl)\, *}_c
\over \psi^{(\Gl)}_1} \ .
 & \Ij b\cr }$$
These $M$'s and $N$'s may be regarded as the structure constants
of two commutative and associative algebras, dual to one another.
They are two generalizations of the Verlinde formula to which
they both reduce in the self-dual case where $\psi=S$.  Beside
these selfdual cases, the $M$'s are in general not integers but
rather algebraic numbers, belonging to the number field generated
by the $\psi$'s. It is therefore surprising that in contrast, the
$N$'s are still found to be  rational numbers,  and in fact in
most  known examples they are  integers.
More precisely, there exists at least one choice of
the vertex ``1", and if some of the eigenvalues are degenerate,
one choice of the basis $\psi$ such that the $N$'s be
rational integers.  This is an empirical observation for which no general
proof is known to us.  Note that we can identify the graph
(that is the set of vertices and the collection of arrows
associated with any $G_p$) with the set of matrices $\{N_b\}$ and
the action of any $G_p$ defined as the (left) matrix
multiplication. Indeed we have
\eqn\IIja{G_p \, N_a =\sum_b (G_p)_{a b} N_b\,,}
which is obtained inserting the r.h.s. of \Ij{b}\ for
$(N_a)_b^c=N_{ab}^c\,,$
 exploiting also the fact that $G_p$ is diagonalised by the same
eigenvectors,
i.e., $(G_p)_{b c}=\sum_{\Gl \in \Exp } \
\zg_p^{(\Gl)}\psi^{(\Gl)}_b\psi^{(\Gl)\, *}_c \,,$
 and using the first orthogonality relation in \Ik{a}.
Taking in the l.h.s of \IIja\ the identity matrix $N_{\un}$
we see that the $G_p$ belong to the  algebra of $N$  matrices and
are in fact linear combinations of these
$N$ matrices with non negative integral coefficients.
Also it is easy to prove that the $N$ algebra is
$\IZ_n$ graded {\it i.e.} that $N_{a b}^c \not=0$ only if
$\tau(c)=\tau(a)+\tau(b)\ \mod n$ as a consequence of the
invariance of the set $\Exp$ under $\Gs$  and of the
transformation \quar\ of the eigenvectors.
In particular $G_1$ and $G_{n-1}$
coincide   with the  matrices $N_{a_f}\,$
and $N_{a_f^{\vee}}\,$  respectively because of ix) above.
The fact that the adjacency matrices are  found within the
$N$ algebra will be used below.

\medskip
These graphs are relevant for two related problems.
First, they are believed to allow the construction of generalized
height (or RSOS) integrable and critical lattice models.
This is true for $n=2$ \VPun\ and has been verified
 also in some cases of $n=3$ \refs{\DFZun{,}\So},
but it is likely that the previous conditions are
not restrictive enough in general.
On the other hand  these lattice models are related to the previous topic of
$sl(n)$ CFT's insofar as their continuous limit is
 described by the minimal coset above. Again, this is convincingly
demonstrated only in the case of $sl(2)$, whereas the other
cases rely on several empirical evidences.

i) The first of these evidences is that the same set of integrable
weights  of $\slh(n)$ at level $k$ that label the diagonal terms
of a modular invariant partition function \Iea\ also describes
the spectrum of eigenvalues of one of the graphs $\CG$.

\lref\ALZ{D. Altsch\"uler, J. Lacki and Ph. Zaugg, {\it Phys. Lett.}
{\bf 205B} (1988) 281.}
 In fact it appears that not all modular invariant partition functions
of $sl(n)$ WZW  theories may be matched with a graph. For
example, no graph satisfying the conditions above can
have a spectrum $\Exp$ matching the diagonal terms of
the infinite series of modular invariants of $\slh(3)$ for a level $k$
not a multiple of 3, \ALZ\
\eqn\Ikor{Z=\sum_{\Gl\in Q\cap \CP_{++}^{(n+k)}}
|\chi_{_{\scriptstyle{\Gl}}}|^2
+\sum_{\Gl\in \CP_{++}^{(n+k)}\\ Q }\chi_{_{\scriptstyle{\Gl}}}
\chi^*_{_{\scriptstyle{\Gs^{k\tau(\Gl)} \Gl}}}\,, \qquad k \ge 4 \ ,}
where $Q$ is the root lattice of $sl(n)$. Indeed the only diagonal
terms in \Ikor\ come from the first sum, and the set
$Q\cap \CP_{++}^{(n+k)}$ is clearly
not invariant under the action of $\Gs$, in contradiction with a
property of the sets $\Exp$, see above.
Another counterexample is provided by the conformal embedding
of $\widehat{A_8}$ into $\widehat{E_8}$, both at level 1.
There is only one
\rep\ of $\widehat{E_8}$ of level one, that decomposes into three
\rep s of
$\widehat{A_8}$, {\it v.i.z.} $\Gr$, $\Gl_3=\Gr+\hat \GL_3$ and
$\Gl_6=\Gr+\hat\GL_6\,$ \KS.  The partition function reads
\eqn\IeAE{ Z=|\chi_{_{\Gr}} +\chi_{_{\Gl_3}} +\chi_{_{\Gl_6}}|^2\ .}
As there are only three contributing \rep s, a possible graph satisfying the
axioms of sect. 2.2 must have three vertices and cannot satisfy \Ig\ for $p=1$.

As far as we  can see, there are two possibilities at this stage.
 One may try to relax some of the conditions that we
have put on the graphs. For example, demanding that
\Ig\ holds in a weaker form, with mod $n$ replaced in the r.h.s.
by mod~$m$, where $m$ is a divisor of $n$,
would allow to accomodate the case \IeAE\ above, in which
$G_1= G_4= G_7\,, \, G_2= G_5= G_8\,, \,
G_3= G_6= \un
 \,,$ and $m=3$. Accordingly the set of
exponents is invariant under the subgroup $\IZ_3=\{1\,, \,\zs^3\,,\,
\zs^6\}\,$ of $\IZ_9\,.$
We shall not pursue this line here. Alternatively,
we may decide to concentrate  on those modular invariants that
are relevant for the coset theories, see \Ieaaa. It is known
\KW\ that the only branching functions (or characters)
 of cosets $G\times G/G$ that do not vanish are those
for which, with our notations of sect. 2.1,
$\Gl-\Gl'-\Gl''+\Gr$ belongs to the root lattice $Q$. For a
modular invariant of the form \Ieaaa, this implies that $\Gl-\bar\Gl$
has to belong to $Q$, which is not the case for the two counterexamples
above. This is this attitude, which is consistent with the general
belief that the graphs are deeply connected with the lattice models,
hence with the coset theories, that was adopted in \DFZ.

ii) The second evidence was pointed out recently in the case
of $sl(2)$ theories \PZ. If one computes the {\it ratios}
of structure constants of spin zero fields in two
theories of identical level, say for the $D$ or $E$ type
of theory over the $A$ theory of same level, one finds
that
\eqn\Ika{M_{\Gl\Gm\Gn}
={D^{(D)\ {\rm or }\ (E)}_
{(\Gl\Gl)(\Gm\Gm)(\Gn\Gn)}\over  D^{(A)}_{(\Gl\Gl)(\Gm\Gm)(\Gn\Gn)} }\ .}
(The reader is referred to \PZ\ for a precise definition
of the constants $D$).

iii) The last evidence was presented in  \DFZ.
There it was shown that for a given graph and the corresponding
modular invariant identified through the observation i) above,
there is an empirical coincidence
between the non-negativity
of the $M_{\Gl\Gm}^{\ \ \Gn}$ and $N_{ab}^{\ \ c}$
coefficients and the property of the corresponding modular
invariant partition function to belong to type I defined in  \Ieb.
Then it was shown that for those graphs that have
the structure constants $N$ and $M$ non negative,
there exists an algebraic way to determine the partition
function of the corresponding CFT.
This was based on the theory of ``$C$-algebras'' that will be
reviewed and used in sect. 5 of the present paper.

In the present paper we want to extend the discussion of the second
point to $sl(n)$ theories, $n>2$ and show its relations with
the last point.


\newsec{ Duality equations and constraints on the  $M$  algebra. }
\nind
Let us consider one
of the euclidean  $4$-point functions
 $$\la \zF_{\zL_1^*}(x_1)\, \zF_{\zL_2^*}(x_2)\, \zF_{\zL_1}(x_3)\,
\zF_{\zL_2}(x_4)\,   \ra\,$$
in the minimal $W_n$ theories \FL.
Here as explained above
$\zL_i=(\zl_i,\bl_i)\,,$ with  $\Gl_i,\bar\Gl_i \in \CP_{++}^{(k+n)} $,
label fields of
the subalgebra of ``thermal" fields.
The label should furthermore include some
index to distinguish fields with the same pair of
$sl(n)$ weights, but for the time being we shall exclude  the cases
with such degeneracies.  Denote by
$\triangle_{\zl}=\triangle_{\zl^*}\,,\ \triangle_{\bl}$ the
conformal weights of the
 field  $\Phi_{\GL}$
and hence by $s(\zL):=\triangle_{\zl} -
\triangle_{\bl} \in \IZ$ its  spin.
The  physical fields   decompose into chiral vertex operators
(see, e.g.\MS)
\eqn\cvo{
\zF_{\zL_1}(z,\bar z)\,|\zL_2 \ra = \sum_{\zL_5\atop \a, \ba}\,
d_{\zL_1 \zL_2}^{\zL_5; (\a,\ba) }\,
{\zl_5\, \,
\choose \zl_1 \zl_2}_{\a}(z) \,
\otimes
\, {\bl_5\, \, \choose \bl_1
\bl_2}_{\ba}(\bar z) \,|\zL_2\ra\,, }
with structure constants  $d_{\zL_1 \zL_2}^{\zL_5; A }\,$ $(
A=(\a,\ba)\,$), $d_{\zL \un }^{ \zL} =1\,.$
The index $\a=1,2, ..., N_{\zl_1 \zl_2}^{\zl_5}\,,$ labels a basis
in the space $V_{\zl_1 \zl_2}^{\zl_5}$ of chiral vertex operators.
 To simplify the notation  this index is omitted whenever
it takes only one value, e.g.,
$N_{\zl  \zl^*}^{\un}=N_{\zl
\un}^{\zl}=1\,$; $N_{\zl_1 \zl_2}^{\zl_3}=$ dim $V_{\zl_1
\zl_2}^{\zl_5} $ is the Verlinde multiplicity \Ie,
\eqn\v{
 N_{\zl_1 \zl_2}^{\zl_3} = N_{\zl_2 \zl_1}^{\zl_3} = N_{\zl_1^*
\zl_2^*}^{\zl_3^*}=N_{\zl_1 \zl_3^*}^{\zl_2^*} \,.}

The requirement of locality of the physical  correlation functions leads,
taking into account the braiding properties of the chiral vertex
operators, to a set of  equations for the structure constants
$d_{\zL_1 \zL_2}^{\zL_5; A }\,.$  In particular exchanging the two
middle fields and selecting the contribution of the identity $\zl_6=\bl_6=
\un=\zr \,$ in the intermediate channel we have
\eqn\loc{
\sum_{\zL_5\atop A, B}\,  (-1)^{ s(\zL_1) + s(\zL_5)}\,
d_{\zL_2^*\zL_5}^{\zL_1; B} \, d_{\zL_1 \zL_2}^{\zL_5; A }\,
F^{\zl_1, \zl_2}_{\zl_5, (\b,\a) ; \un}
\, \bar{F}^{\bl_1, \bl_2}_{\bl_5, (\bb,\ba) ; \un} =
 (-1)^{ s(\zL_2)}\,  d_{\zL_1 \un }^{ \zL_1}
 \, d_{\zL_2^* \zL_2}^{\un}
\,.  }
In \loc\ $F$ ($\bar F$) denote particular elements of the left (right)
fusion matrix
$F^{\zl_1,
\zl_2}_{\zl_5, (\b,\a) ; \zl_6, (\c,\d)}\,,$
 namely those for $\lambda_6=\un \,.$
Upon suitable normalisation, such that   the diagonal
structure constants ($\zL_i=(\zl_i, \zl_i)$) are given by
$d_{\zL_1 \zL_2}^{\zL_5; (\a,\ba)}=\zd_{\a,\ba}  \,,$
we can choose, whenever $N_{\zl_1 \zl_ 2}^{ \zl_5}\,\not =0\,,$
\eqn\fm{
 F^{\zl_1, \zl_2}_{\zl_5, (\b,\a) ; \un}
 =  \delta_{\zs_{123}(\b), \a^* }\,  F_{\zl_1 \zl_ 2}^{ \zl_5}\,: =
\delta_{ \zs_{123}(\b), \a^*} \, \sqrt{{D_{\lambda_5}\over
D_{\lambda_1}\, D_{\lambda_2}}} \,.}
Here $D_{\lambda}$ is the quantum dimension, defined by a ratio
of $\slh{(n)}_k\,$ modular matrix elements,
$D_{\lambda}= S_{\lambda \un}/S_{\un \un}\,.$
In \fm\ $\zs_{123}(\b)$ is the image of the basis element in
 $V_{\zl_1 \zl_2}^{\zl_5} $ labelled by $\b$
under the mapping  $\zs_{123} \equiv
\zs_{13}\, \zs_{12} : \, V_{\zm \zg}^{\zl}
\rightarrow \zs_{13} V_{\zl^* \zg}^{\zm^*}  = V_{\zl^* \zm}^{\zg*}\,$
and  $\a^*$ labels an element in $V_{\zl_1^* \zl_2^*}^{\zl_5^*}$
with complex conjugated  matrix elements  (see \MS).
 From \fm\ and the fact that the quantum dimensions form a
representation of the fusion algebra, it follows that
\eqn\norm{
\sum_{\lambda_5, \a , \b}\, \Big(F^{\zl_1, \zl_2}_{\zl_5, (\b,\a) ;
\un}\Big)^2
= {1 \over
D_{\lambda_1}\, D_{\lambda_2}} \, \sum_{\lambda_5, \a }\,
D_{\lambda_5} = {1 \over D_{\lambda_1}\, D_{\lambda_2}} \,
\sum_{\lambda_5 }\, N_{\lambda_1 \,\lambda_2}^{\lambda_5} \,
D_{\lambda_5} =1\,. }

There exists a basis such that  upon complex conjugation
\eqn\cc{
\({d_{\zL_1
\zL_2}^{\zL_5; (\a,\ba) }}\)^*=d_{\zL_1^* \zL_2^*}^{\zL_5^*;
(\a^*,\ba^*) }  \,.}
Choosing $d_{\Lambda^* \Lambda}^{\un}=(-1)^{s(\Lambda)}\,,$
we have furthermore
(from the locality of the 3-point function and the above choice
of $F^{\zl_1, \zl_2}_{\zl_5, (\b,\a) ; \un}$)
\eqn\tp{
d_{\zL_2^*\zL_5}^{\zL_1; B} = (-1)^{s(\Lambda_1) +s(\Lambda_5)}\,
d_{\zL_1^*\zL_2^*}^{\zL_5^*; \zs_{123}(B)}\,, }
and hence \loc\ simplifies to
\eqn\locb{
\sum_{\zL_5}\, \sum_{\a,\ba}\, |d_{\zL_1 \zL_2}^{\zL_5; A }|^2
\,   F_{\zl_1 \zl_ 2}^{ \zl_5}\,   F_{\bl_1 \bl_ 2}^{ \bl_5}\,
=1\,. }

There is still a ``gauge" freedom left
in determining the constants $d^A$ :  a transformation
$P_{\a \a'}\otimes \bar{P}_{\ba \ba'} \,$ by an unitary
 matrix $P_{\a \a'}$ keeps invariant  equation \loc\
and the chosen normalisations.
\foot{ The matrices $P$ may depend
on the triplet of weights, i.e.,
$P=P_{\zl_1 \zl_ 2}^{ \zl_5}\, .$ The general duality equations are
covariant when transforming both  $d$ and $F$.  Furthermore they
are invariant  under the change $d_{\zL_1 \zL_ 2}^{ \zL_5; A}
\rightarrow \zm_{\zL_1} \zm_{\zL_2} \bar{\zm}_{\zL_5} \,d_{\zL_1
\zL_ 2}^{ \zL_5; A}\,, $ $\, |\zm_{\zL}|=1\,,$ which  preserves
the $2$-point function normalisation.}
In the $sl(2)$ case the constants $d_{(\zl,\zl)\,(\zm,\zm)}^{(\zn,\zn)}\,$
coincide with the relative structure constants represented by the
r.h.s. of \Ika.

We shall make now, following the idea in \BYZ,  two assumptions
which were justified in the
$sl(2)\,$ case \PZ\ by the explicit solutions of the duality equations.

Consider a theory described by a modular invariant partition
function of type I \Ieb, and denote alternatively by $\{\zl \}$
the representation ${\cB}_i$ of the extended algebra, if
$\zl$ appears in its decomposition with respect to $\slh{(n)}_k\,;$
 recall that the
 dimensions $\triangle_{\zl}\,$ for
$\zl \in \{\zl\}\,$ differ by integers.
Each   field operator in these cases is described by a pair of
weights $(\zl, \bl)$ such that both  $\zl, \bl \in \{\zl\}\,.$
 We shall make a frequent use of this feature, carrying out the
summation over pairs of weights (like $\GL_5$ in \locb) in two steps:
a summation over blocks and a summation over weights within that block.
We shall assume that in all nondiagonal cases associated with
conformal embeddings the constants $d^A\,$ factorise according to
\eqn\fa{
N_{\{\zl_1\}, \{\zl_2\}}^{\{\zl_5\}}\,
\sum_{\a,\ba}\, |d_{\zL_1 \zL_2}^{\zL_5; A }|^2 =
\M_{\zl_1 \zl_2}^{\zl_5 }\,\M_{\bl_1 \bl_2}^{\bl_5 }\,,}
where $N_{\{\zl_1\} \{\zl_2\}}^{\{\zl_5\}}$ are the Verlinde
multiplicities of the extended theory,  $\zl_i, \bl_i \in
\{\zl_i\}\,,$  and $\M_{\zl_1 \zl_2}^{\zl_5
}\,$  are some real constants.
They are defined only whenever  $N_{\zl_1 \zl_2}^{\zl_5 }\not =0\,,$
$N_{\{\zl_1\} \{\zl_2\}}^{\{\zl_5\}}\not =0\,,$
and will be considered to be identically zero otherwise. The
above formula,  which allows to block-diagonalise \locb,   makes sense in
the diagonal case as well. Indeed in that
case $N_{\{\zl_1\} \{\zl_2\}}^{\{\zl_5\}}\,$ coincides with the
Verlinde multiplicity and \fa\
reduces to $ \M_{\zl_1 \zl_2}^{\zl_5 }\, = N_{\zl_1\zl_2}^{\zl_5 }\,. $
In the nondiagonal cases
related to conformal embeddings of $\slh(n)_k\,$ into a simple
affine KM algebra $\ggh\,$ at level $1$ the multiplicities $N_{\{\zl_1\}
\{\zl_2\}}^{\{\zl_5\}}$  take the values $1,0$ in all cases.

We insert \fa\ into \locb\ and make now the second assumption
that the quantity
\eqn\class{
N_{\{\zl_1\} \{\zl_2\}}^{\{\zl_5\}}\, F_{\{\zl_1\}
\{\zl_2\}}^{\{\zl_5\}}:= \sum_{\zl_5 \in \{ \zl_5 \}}\, \M_{\zl_1
\zl_2}^{\zl_5 }\,    F_{\zl_1 \zl_2}^{\zl_5} }
depends on $\zl_1\in \{\zl_1\}\,, \, \zl_2 \in \{\zl_2\}\, $ only
through the equivalence classes $\{\zl_1\}\,, \, \{\zl_2\}\,. $

With these assumptions  eq. \locb\ implies
\eqn\normb{
\sum_{\{\zl_5\}} \, N_{\{\zl_1\} \{\zl_2\}}^{\{\zl_5\}}\, \Big(
F_{\{\zl_1\} \{\zl_2\}}^{\{\zl_5\}} \Big)^2 =1\,. }

To be a function of the classes $\{\zl_1\}\, \{\zl_2\}\,,$ $
F_{\{\zl_1\} \{\zl_2\}}^{\{\zl_5\}}\,$ has to be representable in
all possible ways by the $\M$'s according to \class.  Thus given
$ F_{\{\zl_1\}
\{\zl_2\}}^{\{\zl_5\}}\,,$    satisfying \normb, we get a  linear system of
equations for the unknown $\M_{\zl_1 \zl_2}^{\zl_5 }\, $.

 The factor  $ F_{\{\zl_1\} \{\zl_2\}}^{\{\zl_5\}}\,$
in the l.h.s. of \class\
can be interpreted as the counterpart in the extended theory of the
particular fusion matrix elements in \loc\ and \fm;
accordingly we can choose
\eqn\efm{
F_{\{\zl_1\} \{\zl_2\}}^{\{\zl_5\}}=
 \sqrt{{D_{\{\lambda_5\}}\over D_{\{\lambda_1\}}\,
D_{\{\lambda_2\}}}} \,,}
where $D_{\{\lambda\}}= S_{\{\lambda \} \{\un \}}/S_{\{\un\}
\{\un \}}$ is the extended quantum dimension expressed by the
extended modular matrix $ S_{\{\lambda \} \{\mu \}}$.
The set of equations \normb\ follows from the requirement of
locality of the
diagonal correlators of the extended theory and \class\
can be seen as a consistency condition for \normb\ to hold.

This kind of argumentation can be extended as discussed in
the third appendix of reference \PZ.
Namely starting from the general system of equations implied by
locality one can recover the full fusion matrix of the extended
theory and convert the initial  duality equations to  the extended
ones.
  Accordingly a more general  factorisation assumption,
replacing \fa, can be made for the constants
$d^A$ themselves before a summation over $A$. However the
resulting set of equations for the chiral ``halves" of $d^A$ ,
which would replace
\class, involves the unknown general fusion matrix elements
$F^{\zl_1,
\zl_2}_{\zl_5, (\a,\b) ; \zl_6, (\c,\d)}\,.$
Thus, although the full
extended fusion matrices are simple and  in principle known
 (recall that the level of the extended algebra is $k=1$),
this system cannot be effectively used to find the unknown structure
constants $d$, as has been done in  the rank $1$ case
(see Appendix C of \PZ).
Our strategy in what follows will be to study
instead the simpler equations \class\ for the
matrices $\M$. The knowledge of the latter provides partial
information about the structure constants $d$ and completely
determines them  whenever the corresponding Verlinde multiplicity
is equal to $1$.

\medskip\nind
{\bf Remarks}
\item{1)}
 As suggested by the example
 $\slh(2)_4 \subset \slh(3)_1\,$ (i.e., the ``$D_4$'' case),
in which there is a degeneracy of fields, we cannot
expect the factorisation \fa\ to always hold in the ``unitary basis" , i.e.,
for $d^A$ satisfying  \cc.  Indeed, in that case as well as for
 the whole  $\De$ (sub)series  $k=4$ mod $8$,  the basis for
which the relative structure constants are real (recall that
$\zL^*=\zL$ in the $sl(2)$ case) excludes the appearance of more
than one of the ``doubled" scalar fields in any product of fields
and the $d$'s do not factorise.
Complex linear combinations of these real scalar
fields  yields a pair of mutually conjugate fields and complex
structure constants.  The latter  satisfy a generalisation of
\cc, which takes into account the hermitian
conjugation of the two new scalar fields.  Only in this basis the
factorisation  \fa\ takes place. Assuming that similar
considerations extend to higher rank we shall use \class\
also in cases with degeneracies of the exponents. With a slight
 abuse of notation we continue to use $\{ \zl\}\,$ for the
different representations of the extended algebra containing
the $\slh(n)_k\,$
representation $\zl\,$. However in all concrete examples a clear
distinction will be made, adding an additional index to $\zl\,$
so that no confusion arises. The same will be done when
 $\zl$ appears with a nontrivial multiplicity
${\rm mult}_{\{\zl\}}(\zl)\,$ in a given representation $\{\zl\}$;
in expressions like the r.h.s. of \class\ it will be assumed that
the summation runs over all such exponents, i.e.,
$\sum_{\zl \in \{\zl\}} (...) = \sum_{\zl\in  {\CP}^{(h)}_{++}}\,{\rm
mult}_{\{\zl\}}(\zl)\,  (...)\,,$ thus accounting for
the multiplicities. In fact in most of what follows (with the exception
of Appendix A and sect. 5.4) we shall restrict
ourselves to the cases with
trivial multiplicities,  ${\rm mult}_{\{\zl\}}(\zl)=0,1\,.$
\item{2)}
 We recall that in an appropriate basis the factorisation  \fa\
holds  true also in the $sl(2)$ block-diagonal $\De$ series -- whenever  the
structure constants involve the fixed points of \Iaa. Furthermore
the remaining constants factorise similarly to \fa,  but in
a weaker sense, up to a nonchiral $\zd$ -- function factor in the r.h.s.,
due to an intrinsic   $\IZ_2$  grading of the
2-dimensional OPA \PZ;  also the extended
multiplicity  does not appear in the l.h.s. of \fa. Although  a
block-diagonalisation of
\loc\ of the kind  performed  above
is impossible in general,  \normb\ can be effectively recovered,
since the (chiral) equation \class\ is  still valid. The
only difference is that  the extended multiplicity $N_{\{\zl_1\}
\{\zl_2\}}^{\{\zl_5\}}$ in \class\ and \normb\ can take also the value  $2\,.$

\medskip
\bigskip

Combining \fm\ and \efm\ we rewrite \class\
\eqn\sm{
\encadremath{
N_{\{\zl_1\} \{\zl_2\}}^{\{\zl_3\}}  =\sqrt{D_{\{\lambda_1\}}
\over D_{\lambda_1} } \,
\sqrt{D_{\{\lambda_2\}} \over D_{\lambda_2} }\,
\sum_{\zl_3 \in \{ \zl_3 \}}\, \M_{\zl_1 \zl_2}^{\zl_3 }\,
\sqrt{ D_{\lambda_3}\over D_{\{\lambda_3\}}} \,,
\qquad\eqalign{\zl_1&\in \{\zl_1\} \cr
\zl_2 &\in \{\zl_2\}\,\cr }\ .} }
This system of equations
\sm\ that has been derived for the thermal fields of $W_n$ coset
CFT can be analogously  derived
for the $\slh{(n)}_k\,$ WZW models.
\medskip

The constants $\M$ will soon be shown to be identical with the
structure constants $M$ \Ij{a}\ of the Pasquier algebra.
Apparently the input in \sm\ is very simple  -- we need the fusion
rules of the extended theory and the quantum dimensions of  both
the extended and unextended theories, all of which are known,
being expressed by the modular matrices of the corresponding WZW
models. The unextended fusion rules are also implicitly taken
into account since $\M_{\zl_1 \zl_2}^{\zl_3 } \equiv 0\,$ if
$N_{\zl_1 \zl_2}^{\zl_3 } =0\,$.  In fact  this simplicity of
(11) limits its applicability since in general the set of
independent equations in (11) is not sufficient to determine completely
the constants $\M_{\zl_1 \zl_2}^{\zl_3 }\,$ for all values of the
couplings $\big({\zl_1}\,, \, {\zl_2}\,,\,{\zl_3 }\big)\,;$
rather we get some relations between these constants.  Yet there
are some examples, to be discussed in the next section,  in which
the above data provide a full solution.

As suggested by the $\De\,$  $sl(2)$ case, the formula
\sm\  might actually apply to all type I theories,
 though related in a more subtle
 way  to the duality equations of the CFT,  since some substitute
 of \fa\ is required. We check and
confirm this conjecture in Appendix B  for the block-diagonal
$sl(3)\,$ modular invariants of the $D$ series, comparing with the
results in \Ko, \DFZ.

 \medskip


\newsec{Examples of solutions of  \sm.}

\nind
Let us illustrate the previous considerations on the case of
non trivial $\slh{(n)}_k$ theories obtained by conformal embeddings
 into a simple affine algebra $\ggh$. The latter
 have been classified \BBSW, see also \refs{\KS{,}\KW}.
The algebra $\slh{(n)}$ may
be conformally embedded into a few exceptional algebras,
 necessarily of level 1, and
there are (restricting to embeddings into simple algebras) four
infinite series of conformal embeddings
\eqn\IIIa{\eqalign{
\slh{(n)}_{n-2} & \subset \slh{\( {n(n-1)\over 2}\)}_1\,, \qquad
n \ge 4\,, \cr
\slh{(n)}_{n+2} & \subset \slh{\( { n(n+1)\over 2}\)}_1\,, \cr
\slh{(2n)}_{2n} &\subset \widehat{so}(4n^2-1)_1\,, \qquad
n \ge 2\,,  \cr
\slh{(2n+1)}_{2n+1} & \subset \widehat{so}(4n(n+1))_1\,.  \cr}}

We shall look for solutions $\M_{\zl_1 \zl_2}^{\zl_3}\,$
of the system of equations \sm,  which
satisfy the symmetry relations
\eqn\sym{\eqalign{
\M_{\zl_1 \zl_2}^{\zl_3} =&\M_{\zl_2 \zl_1}^{\zl_3} = \M_{\zl_1^*
\zl_2^*}^{\zl_3^*}=\M_{\zl_1 \zl_3^*}^{\zl_2^*} \,,
\cr
&\M_{\zl \zl^*}^{\zr}=1= \M_{\zl \zr}^{\zl}\,,
\cr }}
in agreement with the symmetry of the structure constants $d$ and \v.
 All solutions that we will find are consistent with
 the symmetry under the  $\IZ_n$ automorphism \Iaa
\eqn\aut{
 \M_{\zl_1 \zs(\zl_2)}^{\zs(\zl_3)} =
 \M_{\zl_1 \zl_2}^{\zl_3} \ .}

 Recall that there is a general and explicit formula for
the quantum dimensions in the $\slh(n)_k$ affine algebra in terms of
scalar products with positive roots,
$D_{\zl}= \prod_{\za
>0 }\, {{\rm sin}\big(( \zl, \za ) \pi/(k+n)\big) \over {\rm
sin}\big(( \zr, \za ) \pi/(k+n)\big) }\,.$
 They satisfy various
properties like $D_{\zl}=D_{\zl^*}=D_{\zs(\zl)}$ and $D_{\zl} >0
$ for integrable representations.
The case $sl(2)$ was considered in \PZ. The system \sm\ is
sufficient to recover all $\M \equiv M$ for the block-diagonal
invariants labelled $D_4\,$ ($k=4$) and $E_6\,$ ($k=10$), while
the case $E_8$ ($k=28$) requires considering a system of equations
larger than \sm, as discussed above.

\medskip
Let turn now to the case $sl(3)$. There  are four values of
the level, $k=3\,, 5\,, 9\,, 21\,, $ for which there exist
modular invariants originating from some diagonal modular
invariant of the  embedding algebra.
The $\slh(3)$ modular invariant for $k=3$ corresponds to the embedding
$\slh{(3)}_3 \subset
\soh(8)_1\,,$  and it is also one of the invariants
of the $\CD$ series for $k=0$ mod $3\,,$ \B\
\eqn\inv{
Z_{\slh{(3)}_3}=
|\chi_{_{11}}+\chi_{_{41}}+\chi_{_{14}}|^2+ 3 |\chi_{_{22}}^2|\,.}
The integrable representations of $\soh(8)_1\,,$ are
given by the identity
 $(1,1,1,1)$ and $(2,1,1,1)\,,$ $(1,1,2,1)\,,$  $(1,1,1,2)$
(with standard notations for the $so(n)$ weights).
All have quantum dimensions
equal to $1$ and the charge conjugation is trivial. Their fusion
algebra is of type ${\cal D}_2 \cong \IZ_2 \times \IZ_2$.
\eqn\f{\eqalign{
&w_i \times w_i = \un\,, \quad i=1,2,3\,,
\cr
&w_i \times w_j =w_k\,, \quad i\not = j\not = k \not =i\,.
\cr}}
The three identical  representations $(2,2) \,$ of $\slh{(3)}_3\,$
are distinguished naturally by assigning  to each of them
the index $i$ inherited from the $\soh(8)_1\,$ counterpart.
The only nontrivial $\slh{(3)}_3\,$ quantum dimension is
$D_{2,2}=3\,.$ The system reduces,  after taking into account \sym, to
\eqn\smb{\eqalign{
&1= \M_{(2,2)_i \ (4,1) }^{(2,2)_{i} }\,,\qquad 1 =\M_{(4,1) \
(4,1)}^{(1,4)}\,,
\cr
&1= \M_{(2,2)_i \ (2,2)_j }^{(2,2)_{k} }\ {1\over
\sqrt{D_{(2,2)}} }\,, \quad
{\rm for} \quad i\not = j\not = k \not =i\,,
\cr
&1= \(\M_{(2,2)_i \ (2,2)_i }^{(1,1)} +\M_{(2,2)_i \ (2,2)_i
}^{(4,1)}+\M_{(2,2)_i \ (2,2)_i }^{(1,4)}\) \ {1\over D_{(2,2)}}\,,
\cr}}
which implies using \sym\ again
\eqn\vmb{\eqalign{
&\M_{(2,2)_i \ (2,2)_j }^{(2,2)_{k}}= \sqrt{3}\,, \quad {\rm for}
\quad i\not = j\not = k \not =i\,,
\cr
&\M_{(2,2)_i \ (2,2)_i }^{(1,1)} =\M_{(2,2)_i \ (2,2)_i
}^{(4,1)} = \M_{(2,2)_i \ (2,2)_i }^{(1,4)} =1\,.
\cr}}
The last two equalities in \smb\ incorporate the $\slh{(3)}_3\,$
fusion rule
$$
(2,2) \otimes (2,2) = (1,1)\oplus(1,4)\oplus(4,1)\oplus 2 (2,2)
$$
in a way respecting the fusion rule \f\ of the corresponding
extended representations.
The remaining  matrix elements are recovered from \sym;  \aut\
is indeed fulfilled.

In this and in all examples to follow  the degeneracy
of fields with the same $sl(n)$ weight is resolved
by the correspondence to different representations of the extended  algebra.
In all these cases the symmetry
relations have to be reformulated for the pairs of indices,
taking into account also the conjugation properties
(trivial in the above example) of the extended fusion
multiplicities.

 To the modular invariant \inv\ has been assigned
in \Ko\DFZ\ the graph $\CD^{(6)}$ of Fig. 1.
For an appropriate choice of the
eigenvectors $\psi$, the $M$'s computed according to \Ij{a}
coincide with the $\M$ just determined.
 This coincidence can be also checked on the example $k=5\,.$
The WZW exceptional modular invariant described by the conformal
embedding
$\slh{(3)}_{5} \subset \slh{(6)}_{1} $  reads \ChR
\eqn\invb{
Z_{\slh{(3)}_5}=
|\chi_{_{11}}+\chi_{_{33}}|^2+|\chi_{_{13}}+\chi_{_{43}}|^2 +
|\chi_{_{23}}+\chi_{_{61}}|^2 + |\chi_{_{41}}+\chi_{_{14}}|^2
+|\chi_{_{32}}+\chi_{_{16}}|^2 + |\chi_{_{31}}+\chi_{_{34}}|^2 \,.}
 The integrable representations  of $\slh{(n)}_{1}$ consist
 of the identity $\zr=\sum_{i=1}^{n-1} \, \GLh_i\,,$ and the
fundamental representations $
\GLh_1+\zr =\zs(\zr)\ ,\cdots\,,
\, \GLh_{n-1}+\zr = \zs^{n-1}(\zr)\,.$ They close on a $\IZ_n$ type
of fusion algebra, with
$\Gl_i:=\GLh_i+\zr\,, $  being identified with $\zs^i\,,$
and  have quantum dimensions equal to $1$.
 In our example these representations correspond to the blocks
in the above order, i.e.,
 $\chi_{_{13}}+\chi_{_{43}} \rightarrow \zs\,,$
  $\chi_{_{23}}+\chi_{_{61}} \rightarrow \zs^2\,,$
etc. The  quantum
dimensions of the relevant representations of $\slh{(3)}_5\,$ read
\eqn\qd{\eqalign{
&D_{3,3}=D_{3,2}= (1+\sqrt{2})^2 =3+2 \sqrt{2}\,,
\cr
&D_{1,3}=D_{1,4}= D_{3,4} =2+\sqrt{2}\,,
\cr
&D_{1,1}=D_{1,6} =1\,.
\cr }}
Now we can solve the system \sm, imposing the restrictions
of the $\IZ_n$ fusion algebra of the classes on the corresponding
representatives, while the sum in the r.h.s. goes over
representations $\zl_3$
of  $\slh{(3)}_5\,,$ $\zl_3 \in \{\zl_3\}$, such that
 the Verlinde multiplicities $N_{\zl_1 \zl_2}^{\zl_3 } \,$ are nonzero.
Taking into account the  symmetry relations \sym, \aut, the result for the
independent nonzero matrix elements is
\eqnn\vm
$$\eqalignno{
&\M_{(3,3) (3,3)}^{(3,3)}=\sqrt{D_{3,3}} - {1 \over
\sqrt{D_{3,3}}}=2\,,
\cr
&\M_{(3,1)\ (4,3)}^{(3,3)} ={D_{(1,3)} \over \sqrt{D_{3,3}}}=
\sqrt{2}  \,, & \vm
\cr
&\M_{(3,3)\ (3,1)}^{(3,1)}=1=\M_{(3,3)\
(3,2)}^{(1,6)}=\M_{(3,3)\ (4,3)}^{(4,3)}
=\M_{(6,1)\ (3,1)}^{(4,3)}
= \M_{(6,1)\ (1,3)}^{(4,1)}
\,,   \cr}$$
which is in agreement with the values  for the corresponding
matrices $M$  attached to the graph $\CE^{(8)}$ introduced in \DFZ\
(see Fig. 1).
%
%
%
\fig{The oriented graphs $\CD^{(6)}$ and $\CE^{(8)}$ of matrix
$G_1$ for  the conformal embeddings of $\slh(3)_3\subset
\soh(8)_1$ and $\slh(3)_5\subset \slh(6)_1$}
{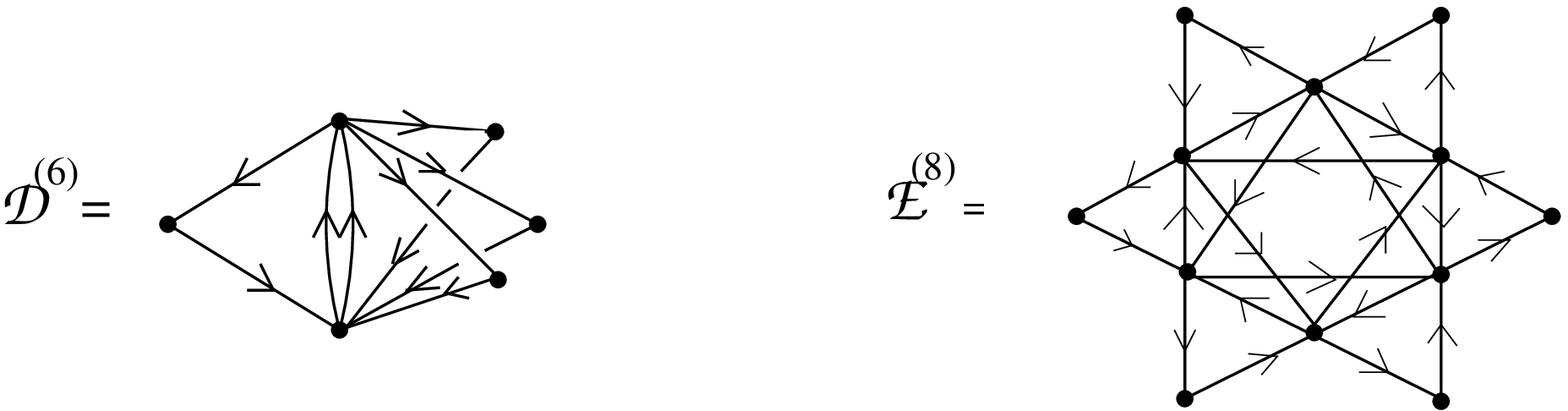}{8cm}\figlabel\graf

The system \sm\ can be analysed also in the remaining two exceptional
cases $k=9,
21\,,$ corresponding to
the embeddings of $\slh{(3)} $ into level $1$ $\widehat{E_6}\,$
and $\widehat{E_7}\,$ respectively. Again all quantum dimensions
of the representations of the extended algebras are equal to $1$, while their
fusion algebras are of type $\Z_3$ and $\IZ_2$ respectively.
However the equations in these cases are
 not sufficient to determine all matrix
elements. Nevertheless they impose a set of restrictions, which
can be checked to be consistent with the explicit results of \DFZ.
As a side remark, we note that  in  the case $k=9$ some of
the squares of the $M$ matrix elements are {\it not} given by rational
numbers. Recall that  they are rational  in the
$sl(2)$ case as a general property \PZ, \RST.

Based on the results in the $sl(2)$ and $sl(3)$ cases
we now make the conjecture that the matrices $M$ and $\M$
introduced in \Ij{a} and \fa\ are identical in general.
Hence we look for new solutions of \sm, to be used to reconstruct
the corresponding graphs as explained in sect. 1.

The first two non trivial cases of the first series in \IIIa\ are
$\slh{(4)}_2\subset \slh{(6)}_1$ and
$\slh{(5)}_3\subset \slh{(10)}_1$,
for which the modular invariant partition functions read respectively
\eqna\IIIb
$$\eqalignno{
Z_{\slh{(4)}_2}&= |\chi_{_{111}}+\chi_{_{131}}|^2+2|\chi_{_{121}}|^2
+2|\chi_{_{212}}|^2+|\chi_{_{113}}+\chi_{_{311}}|^2\,, & \IIIb a\cr
Z_{\slh{(5)}_3}&=|\chi_{_{1111}}+\chi_{_{1221}}|^2+|\chi_{_{3121}}+
\chi_{_{1213}}|^2
&\IIIb b \cr
+ \Big\{|\chi_{_{1211}}&
+\chi_{_{2131}}|^2+|\chi_{_{1411}}+\chi_{_{1212}}|^2
+|\chi_{_{1311}}+\chi_{_{2113}}|^2 +|\chi_{_{4111}}+\chi_{_{2122}}|^2
+ {\rm c.c.}\Big\}\ . \cr}$$
 In the case \IIIb{a}, the successive blocks correspond respectively
to the weights $\Gl_0=\Gr$, $\Gl_1$ and $\Gl_5$, $\Gl_2$ and $\Gl_4$,
and $\Gl_3$ of $\slh(6)_1\,$, where we are making use of the notation
$\Gl_i=\Gr+\GLh_i\,$ introduced above for the fundamental weights shifted by
$\Gr$. In the case \IIIb{b} likewise, the blocks correspond to
$\Gl_0$, $\Gl_5$,  followed by $\Gl_1$,  $\Gl_2$, $\Gl_3$,  $\Gl_4$ and
their conjugate.
 For each of these cases, the system \sm\ fully determines the $\M$'s.
In the case of $\slh{(4)}_2$, the degeneracy of, say, the
two representations $\Gl=(1,2,1)$ is lifted by the assignment of two
opposite $\IZ_6$ charges equal $\pm 1$ that they inherit from the ``parent"
 representations $(2,1,1,1,1)$ and $(1,1,1,1,2)$ in
$\slh{(6)}_1$. The symmetry relations \sym\ hold with,
e.g.,  $((1,2,1)_{\pm})^*= (1,2,1)_{\mp}\,,$ etc.  We thus
obtain for the matrix $\M_{(121)_+}$ in the basis
$(111),\,(131),\,(121)_+,\,(212)_+,\,(311),\,(113),\,(212)_-,\,(121)_-$
$$ \M_{(121)_+}=\pmatrix{ & &1& & & & & \cr
                          & &1& & & & & \cr
                          & & &\sqrt{2}& & & & \cr
                          & & & &1&1& & \cr
                          & & & & & &1& \cr
                          & & & & & &1& \cr
                          & & & & & & &\sqrt{2}\cr
                         1&1& & & & & & \cr} \ .$$
%
%
%
\fig{The graphs of $G_1$ and $G_2$ corresponding to the conformal embedding
$\slh(4)_2\subset \slh(6)_1$.  The involution $a\to a^{\vee}$
acts as the reflection in the line 1--4 if the graph is regarded
as  two dimensional. }
{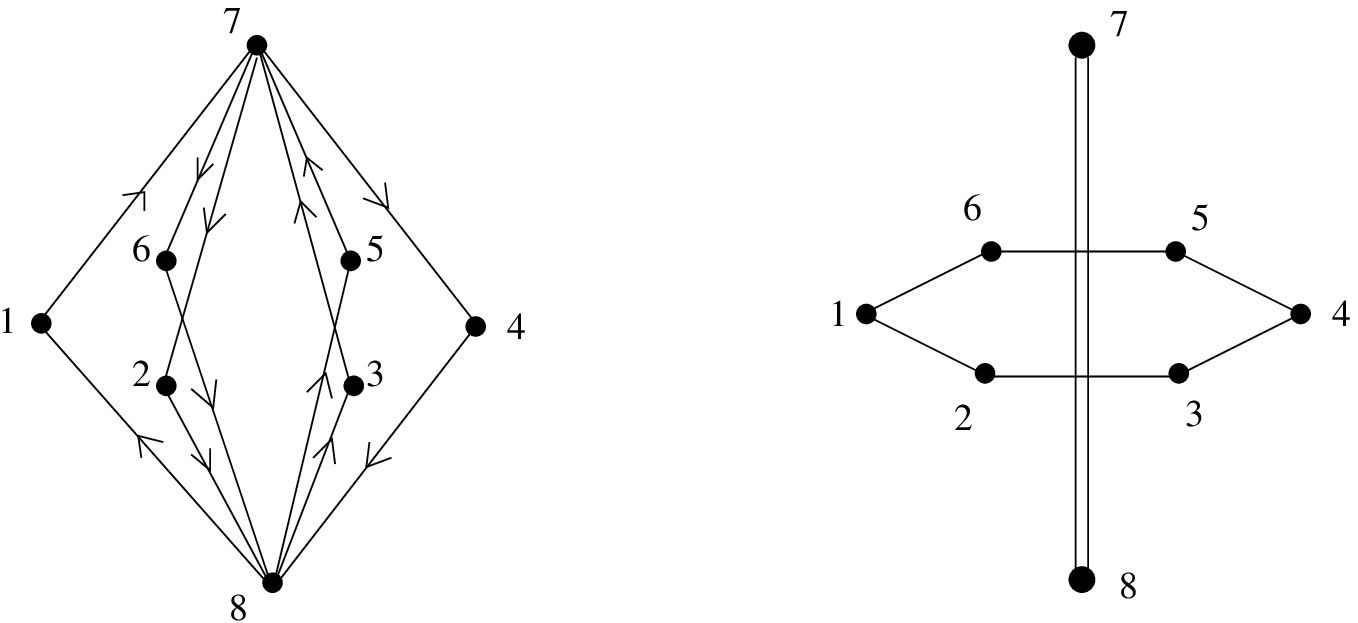}{8cm}\figlabel\diam
%
%
\fig{The graphs of $G_1$ and $G_2$ corresponding to the conformal embedding
$\slh(5)_3\subset \slh(10)_1$; see the text for the conventions
on the $5$-ality of the vertices and the ensuing orientations of edges; for
the graph of $G_2$ a certain reshuffling of the radial directions has been
done for more readability and broken lines indicate double edges.
The involution $a\to a^\vee$
is the reflection in the line 1--6 for  $1\le a\le 10$
whereas $11^\vee=16$ and $a^\vee=32-a$, $12\le a \le 20$.}
{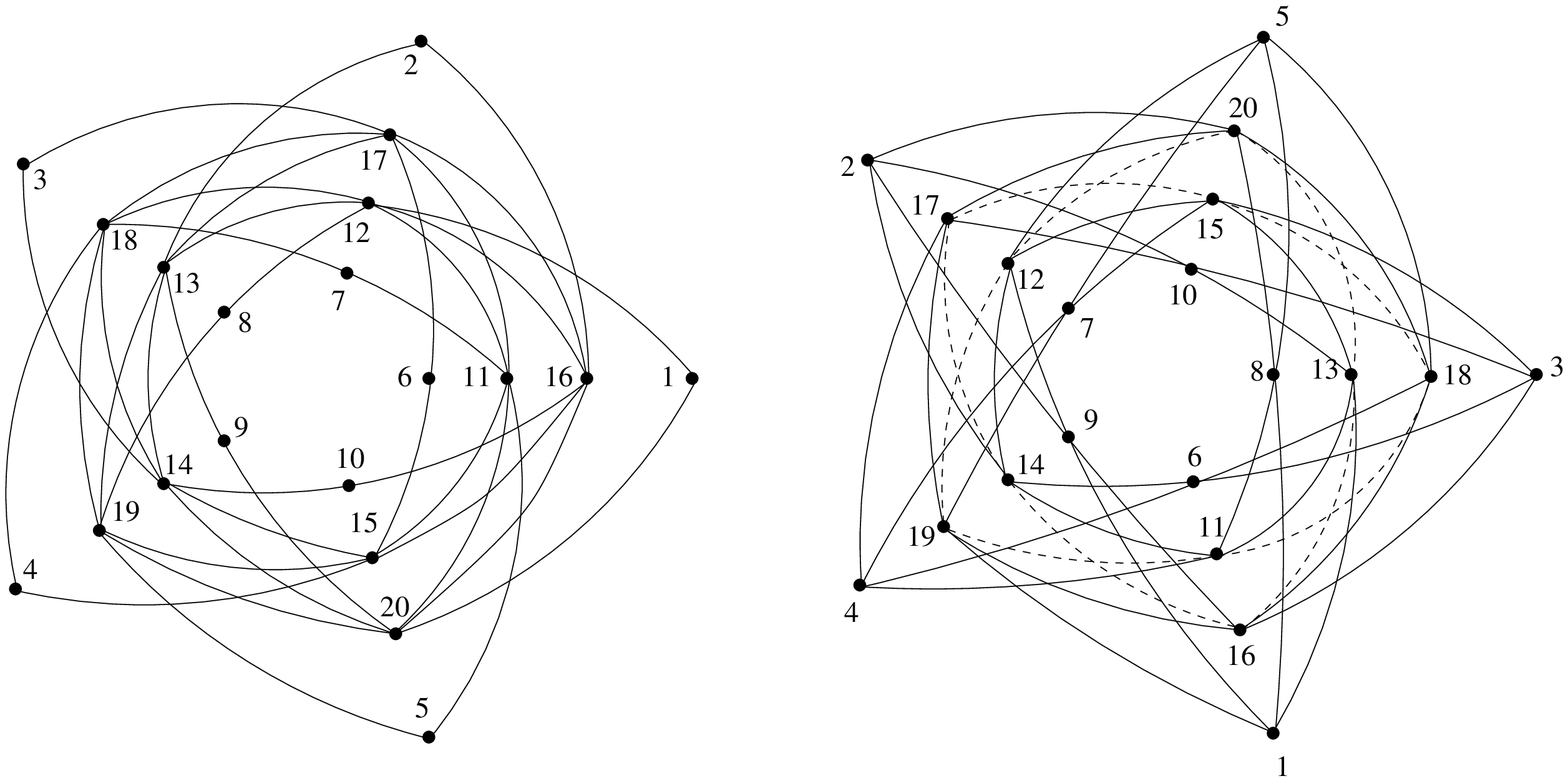}{12cm}
\figlabel\rosebud
Now assuming that $\M=M$ and  diagonalizing it  as in \Ij{a}, we may determine
the $\psi$ and construct the matrices $N$.
As  anticipated
their entries are all non negative integers and one of them,
arbitrarily denoted $N_7$,  turns out to have the desired spectrum to
qualify as $G_1$: its eigenvalues are of the form \Ii\ with $\Gl$
taking all the values that appear in the diagonal terms of \IIIb{a}.
The graph is depicted in Fig. 2 together with the graph of the matrix
$G_2=N_2+N_6$ also identified by its spectrum; finally $G_3=G_1^t=N_8$.
The same procedure with the case of $sl(5)_3$ produces the
graphs of Fig. 3. In fig. 3.a, is represented the graph
of matrix $G_1=N_{12}$, with a certain labelling of the vertices:
this labelling has been chosen in such a way that a vertex
marked $\ell$ had a ``5-ality'' equal to $\ell-1\ \mod 5$.
In fig 3.b, is represented the graph of $G_2=N_{8}+N_{18}$.
 The matrices  $G_3=G_2^t\,, G_4=G_1^t\,$ are also found among the
$N$ matrices or their combinations:
$G_3=N_9+N_{19}$,
$G_4=N_{20}$.
The arrows of these graphs have been omitted as redundant since
they are dictated by rule \Ig.

\medskip


\newsec{
Dual pairs of $C$ - algebras -- properties and applications.}
\nind
In this section we transform  \sm\ into  a  system of equations
for the eigenvectors $\psi_a$ of the $M$ - matrices and
in particular we obtain for  a subset of them  explicit
analytic expressions valid for arbitrary $sl(n)\,.$
 For this purpose
we need some properties of the $C$ (``character") --
algebras  \BI, so let us briefly review the basic notions and
some  theorems involved.

%
\subsec{A short review of the theory of $C$--algebras}
\nind
A $C$ -- algebra $\cU$ is an associative commutative algebra over
$\IC\,,$ with a basis $ x_1\,, x_2\,,...\,, x_d$, real structure
constants,
\eqn\al{
x_a\, x_b = \sum_c \, p_{a b}^c\, x_c\,, \quad p_{a b}^c \in \IR\,,}
an identity $x_1\,,$ (to be  also denoted by $x_{\un}\,$),
hence $p_{1 b}^a= \zd_{a b}\,,$ and an involution $a \rightarrow a^{\vee}\,,$
which extends to an algebra isomorphism $x_a \rightarrow
(x_a)^{\vee} : = x_{a^{\vee}}\,$ --  hence $ p_{a b}^c =
p_{a^{\vee} b^{\vee}}^{c^{\vee}}\,,$ and such that
\eqn\per{
 p_{a b}^1= \zd_{a b^{\vee}}\, k_a\,, \quad k_a >0 \quad \forall
\ \ a,b\,.}
 It is furthermore assumed
that the mapping  $x_a \rightarrow k_a\,$ is a linear
representation of $\cU\,.$ It follows that $\un^{\vee}=\un\,,
\quad k_{\un}=1\,, \quad k_{a^{\vee}}=k_a\,,$ and $k_c\, p_{a
b}^c=k_b\, p_{a^{\vee} c}^b= k_b\, p_{a c^{\vee}}^{b^{\vee}}\,.$
The algebra is realised isomorphically by the matrix commutative
algebra spanned by $(p_a)_b^c\,, \, a=1,.., d.$

These axioms imply furthermore that the matrices $(B_a)_b^c =
{\sqrt{k_c}\over \sqrt{k_b}}\,(p_a)_b^c\,$  are normal
(since $B_a$ commute and  ${}^tB_a=B_{a^{\vee}}\,$)
and hence can be diagonalised simultaneously by an unitary matrix $U$, i.e.,
\eqn\Va{
{\sqrt{k_c}\over \sqrt{k_b}}\, p_{a b}^c =
 \sum_{i=1}^d\, U_{b i}\, p_a(i) \, {U_{c i}}^*\,,}
 where the eigenvalues may be written
$p_a(i) = { U_{a i}\over  U_{1 i}}\, \sqrt{k_a}\,;$   \quad
one may choose $p_a(1)=k_a\,,$  and  the complex conjugate of
$U_{a i}$ satisfies ${U_{a i}}^*=U_{a^{\vee} i}\,.$

The $C$ - algebra  $\cU$  has  exactly $d\,$
linear representations $\triangle_i\,,$
given by $\triangle_i: x_a  \rightarrow p_a(i) \,.$

There is a natural notion of a dual $C$ -- algebra $\hat{\cal U}\,,$
as  the set of all linear mappings of $\cU$ into $\IC$ with
the point-wise multiplication.
It has structure constants $\hat{p}_{i k}^j$ satisfying
\eqn\Vb{
{\sqrt{\hat{k}_k}\over \sqrt{\hat{k}_j}}\, \hat{p}_{i j}^k =
 \sum_{a=0}^d\, U_{a j}\, \hat{p}_i(a) \, {U}^*_{a k}\,,}
 where the eigenvalues are
$\hat{p}_i(a) = { U_{a i}\over  U_{a 1}}\, \sqrt{\hat{k}_i}\,,
\quad \hat{p}_i(1) =\hat{k}_i
\,,$  and ${U_{a i}}^*=U_{a i^{\vee}}\,.$ The dual of
$\hat{\cal U}\,$ is $\cU\,.$

The unitary matrix $U$, which diagonalises the structure constants of
the pair of algebras ${\cal U}\,$ and $\hat{\cal U}\,$ according to
\Va, \Vb\ is in general nonsymmetric.
 Whenever $U$  can be chosen symmetric the algebra $\cU$
is selfdual.

Comparing with \Ij{a}, \Ij{b}, it is clear that we can look at the pair of
algebras spanned by
the $N$ and $M$ - matrices as matrix realisations of
a  dual pair  of $C$ - algebras ${\cal U}\,$ and $\hat{\cal U}\,$ with
\eqna\Vc
$$\eqalignno{
\sqrt{k_a} = {\psi_a^{(\zr)} \over \psi_{\un}^{(\zr)}}\ , &
\quad \sqrt{\hat{k}_{\zl}} = {\psi_{\un}^{(\zl)} \over \psi_{\un}^{(\zr)}}\,,
&\Vc a\cr
N_{a b}^{c}={\sqrt{k_{c}} \over  \sqrt{k_{a}} \sqrt{k_{b}}}\,
p_{a b}^{c}\ ,&  \qquad
 M_{\zl \zm}^{\zg} = {\sqrt{\hat{k}_{\zg}} \over \sqrt{\hat{k}_{\zl}}
\sqrt{\hat{k}_{\zm}}}\, \hat{p}_{\zl \zm}^{\zg}  \,. & \Vc b\cr }
$$

One  introduces in an obvious way the notion of a
$C$--subalgebra of the $C$--algebra $\cU$
as the $C$--algebra   $\cU_T\,$ with basis  $\{x_{a}\,, \, a \in
T \}\,,$ where  $T$ is a subset of $\{1,2,.., d\}\,.$
  $\cU_T$ is a proper $C$--subalgebra if $1 < |T|< d$.

{}From now on we shall assume that both  ${\cal U}\,$ and $\hat{\cal U}\,$
have  nonnegative structure constants.  One  can prove that under
this assumption $\cU$ has a proper $C$--subalgebra iff its dual
$\hat{\cal U}\,$ has a  proper $C$--subalgebra.  (Recall that
the nonnegativity of the $M$ and $N$ matrices was established  for
all known examples of type I theories  and  assumed for all
general theories of this type.) Furthermore if $\cU$ has a proper
$C$--subalgebra   $\cU_T\,$  one can define a factor $C$--subalgebra,
denoted $\,\cU/\cU_T\,,$ by splitting the set
$\{1,2,.., d\}\,$ into
equivalence classes $T_1\equiv T\,, T_2\,, ... , T_t\,.$
Namely $a \sim c $ iff $\exists\,$ $\zb \in
T\,,$ such that $p_{a \zb}^c \not = 0\,.$
 Then the elements $X_i:= {1\over \sum_{a \in T}\, k_{a}}\, \sum_{b
\in T_i}\, x_b\,,$ $i=1,2,...,t\,$, with the multiplication
inherited from $\cU\,$, provide a basis for
$\,\cU/\cU_T\,.$ The parameter $k_{T_i}\, $ is given by
$k_{T_i}= \sum_{a \in T_i}\, k_{a} / \sum_{a \in T}\, k_{a} \,. $
 A theorem in \BI\  states that the dual of
$\,\cU/\cU_T\,$  is isomorphic to a $C$--subalgebra of
$\hat{\cal U}\,,$
obtained by the same procedure
starting from a subset $\hat T$ of the dual basis and thus
denoted $\hat{\cal U}_{\hat T}\,.$
The classes $T_i\,$ can be enumerated by the elements of the
subset $\hat T \ni i \,.$ Vice versa, there is one to one
correspondence between the elements of $T$ and the equivalence
classes $\hat{T}_a\,, a =1,2,...,
|T|\,,$ $\hat{T}_1\equiv \hat{T}\,,$ of $\hat{\cal U}\,,$ and
$\hat{\cal U}/\hat{\cal U}_{\hat
T} \cong \widehat{\cU_T} \,.$

%
\subsec{Further analysis of Eq. \sm}
\nind
Let us  now  return to the set of equations \sm\ assuming that all its
solutions for the  $M=\M$ are nonnegative.
We multiply both sides with
$$
\chi_{\{\zl_3\}} (\{\zg\}):={S_{\{\zl_3\} \{\zg\}} \over S_{\{1\}
\{\zg\} }}\,,
$$
where $S_{\{\zl\} \{\zg\}}\,$ is the modular matrix of the
extended theory. Then we sum
over the classes, using that $\chi_{\{\zl\}} (\{\zg\})$
provide for any $\{\zg\}$ a $1$- dimensional
representation of the extended fusion algebra. We thus obtain
\eqn\Vv{
\Big(\chi_{\{\zl_1\}} (\{\zg\})\ \sqrt{{ D_{\zl_1}\over
D_{\{\zl_1\}}}} \Big)\,\,
\Big(\chi_{\{\zl_2\}} (\{\zg\}) \sqrt{{ D_{\zl_2}\over D_{\{\zl_2\}}}}\Big)=
\sum_{\zl_3}\ M_{\zl_1 \zl_2 }^{\zl_3}
\,\,\Big(\chi_{\{\zl_3\}}(\{\zg\})\,  \sqrt{{
D_{\zl_3}\over D_{\{\zl_3\}}}} \Big)\,.}
It follows that we can interpret $ \chi_{\{\zl\}}(\{\zg\})\,  \sqrt{{
D_\zl\over D_{\{\zl\}}}}$ as some of the $1$ - dimensional
representations of the $M$
algebra, given in general by ${\psi_{c}^{(\zl)} \over
\psi_{c}^{(\un)} }\,,$ $c\in \CV\,.$ Thus identifying each class
$\{\zg \}$ with a vertex  $c$, such that  $\{\un\}$
corresponds to the identity vertex element $\un\in \CV\,,$
and $\{\zg\}^* = c^{\vee}$, we select a subset $T$ of the
vertices for which \Vv\ provides explicit expressions for  the
$M$ -- eigenvalues
\eqn\Vd{
{\psi_{c}^{(\zl)}\over \psi_{c}^{(\un)}}:=
\chi_{\{\zl\}} (\{\zg\})\  \sqrt{{
D_\zl\over D_{\{\zl\}}}} = {S_{ \{\zg\} \{\zl\}} \over S_{
\{\zg\} \{\un\} }} \ \sqrt{{
D_\zl\over D_{\{\zl\}}}}\,, \quad  \forall \Gl\in \Exp\,,
\quad \{\zg \}  \equiv c \in T \,,}
and in particular
\eqn\Ve{
{\psi_{\un}^{(\zl)} \over \psi_{\un}^{(\un)}}=
\chi_{\{\zl\}} (\{\un\})\  \sqrt{{D_\zl\over D_{\{\zl\}}}}=
{S_{ \{\un\} \{\zl\} }\over S_{ \{\un\}
\{\un\} }} \ \sqrt{{ D_\zl\over D_{\{\zl\}}}}=
\sqrt{D_\zl \ D_{\{\zl\}}} \,.}
so that for any exponent $\zl\,,$ we have $\sqrt{\hat{k}_{\zl}}=
{\psi_{\un}^{(\zl)} \over \psi_{\un}^{(\un)}}=\sqrt{D_{\zl}D_{\{\zl\}}}\,.$

 Clearly the l.h.s. of \Vd\ is constant when $\zl$ varies within a
given class $\{\zl\}$
 and hence we can rewrite  \Vd\ more symmetrically dividing by
 $\sqrt{D_{\{\zl\}}}\,$ and summing over $\zl \in \{\zl\}$
$$
\chi_{\{\zl\}} (c) = {\sqrt{D_{\{\zl\}}}\over |\{\zl\}|}\,
\sum_{\zl \in \{\zl\}}\,
   { \chi_{\zl} (c) \over {\sqrt{D_{\zl}}} }\,
\,, \quad c\in T\,, $$
where  $\chi_{\zl} (c)$  stay for the eigenvalues in the  l.h.s. of \Vd\
and $|\{\zl\}|=\sum_{\zl\in \CP_{++}}\, {\rm mult}_{\{\zl\}}(\zl) \,.$

 Using the unitarity relations \Ik{a}\ and the fact that the extended
modular matrix elements $S_{\{\un\} \{\zl\}}\,$ are expressed
through elements of the
modular matrix $S_{\zl \zm}\,$ of $\slh{(n)}_k\,,$    i.e., $S_{\{\un\}
\{\zl\}}\,= \sum_{\zl\in \{\zl\}}\,
S_{\un \zl}\,,$  we obtain from \Vd\
\eqn\Vf{
\big(\psi_{c}^{(\un)}\big)^2=
S_{\un  \un}\ S_{\{\un\}  \{\un\}} \,  D_{\{\zg\}}^2   =
{(S_{\{\zg\}\  \{\un\}})^2 \over \sum_{\za \in \{\un\}} D_\za}
=\big(\psi_{\un}^{(\un)}\big)^2 \,\  D_{\{\zg\}}^2
\,, \quad c\equiv \{\zg \} \in T \,.}
Recall that  (cf.  \Ik{b}\ ) $\psi_{c}^{(\un)}$ are real in
agreement with the reality of $S_{\{\un\} \{\zl\}}\,$ and we can
furthermore choose them positive.  We can thus identify
all the $T$-components of the Perron-Frobenius eigenvector of the
$N$ algebra
\eqn\perron{\sqrt{k_a}= { \psi_{a}^{(\un)} \over
\psi_{\un}^{(\un)}}=D_{\{\za\}}\,, \qquad \forall   a\in T\,.}

Inserting \Vf\ back into \Vd\ we obtain
 an explicit expression for the subset of the eigenvectors
$\{\psi_{c} \,, \, c\in T\, \}$ of the $M$ matrices
\eqn\Vdb{ \encadremath{
{\psi_{c}^{(\zl)}
= S_{\{\zg\}  \{\zl\}}\ \sqrt{S_{\un\, \zl} \over S_{\{\un\}  \{\zl\}}}
=S_{\{\zg\}  \{\zl\}}\ \sqrt{D_{ \zl} \over
\sum_{\zl \in \{\Gl\} }\,   D_{ \zl}} }
\ ,
\qquad  \eqalign{&\forall \Gl\in \Exp\cr& \{\zg \} \equiv c  \in T \cr}\ .}}

The formula  \Vdb\ is the main result in this section.
In particular it determines explicitly the Perron -- Frobenius
vector  (of the $M$ algebra)
\eqn\Vdba{\psi_{1}^{(\zl)}= \sqrt{S_{\un\, \zl}  S_{\{\un\}
\{\zl\}}}\,, \quad \zl \in \{\zl\}\,.}
Both $S_{\un\, \zl}$ and $  S_{\{\un\}
\{\zl\}}$ are real, positive -- and thus taking the positive root
in \Vdb, all components of $\psi_{1}$ are indeed positive. The
formula \Vdb\ furthermore implies
\eqn\Vg{
{ \psi_{c}^{(\zl)} \over \psi_{\un}^{(\zl)}}=
{S_{\{\zg\}  \{\zl\} }\,\over S_{\{\un\}
\{\zl\} }}\, = \chi_{\{\zg\}} (\{\zl\})\,, \quad
\, c\equiv \{\zg\} \in T \,,
 \quad \zl\in \{\zl\}\,, }
(i.e., the eigenvalues of $N_c\,, \, c\in T\,,$ are constant
within a class $\{\zl\}$) and  hence
\eqn\Vh{
{ \psi_{c}^{(\zl)} \over \sqrt{ \sum_{a\in T } \
|\psi_{a}^{(\zl)}|^2}}  = S_{\{\zg\}  \{\zl\} }\,,
 \quad   \, c\equiv \{\zg\} \in T \,,
  \quad \zl\in \{\zl\}\,, }
\eqn\Vi{
\sum_{\zl \in \{\zl\} }\ \sum_{a \in T}\
|\psi_{a}^{(\zl)}|^2 = 1\,,}
since $ \sum_{a\in T } \ |\psi_{a}^{(\zl)}|^2 =
S_{\un\, \zl}/ S_{\{\un\}   \{\zl\}}\,.$
(Note also $ \sum_{\zl \in \{\zl\} } \
 |\psi_{c}^{(\zl)}|^2 = |S_{\{\zg\}\{\zl\}}|^2 \,$  for $ c\equiv
\{\zg\} \in T\,,$ so that $ \sum_{\zl \in \Exp } \
 |\psi_{c}^{(\zl)}|^2 = 1\,,$ in agreement with \Ik{a}.)

 The symmetry of the extended modular
matrix  $S_{\{\zg\}  \{\zl\} }= S_{\{\zl\}  \{\zg\} }\,$ leads
to relations
for the $\psi$'s, which, using \Vdb, read
$$ \sqrt{{D_{\{\zb \} } \over D_\zb }} \  \psi_{a}^{(\zb)}=
\sqrt{{D_{\{\za\} }\over D_\za}} \  \psi_{b}^{(\za)}\,, \quad
 a,b \in T\,, \, a\equiv \{\za\}\,, \, b \equiv \{\zb\}\,, $$
i.e., the ratio of quantum dimensions ``deforms" the nondiagonal
matrix elements $  \psi_{a}^{(\zb)}\,,$ $ a,b \in T\,.$ If each
class contains only one element, i.e., $T$ coincides with $\CV$,
and $D_{\{\zl \} }\equiv D_\zl\,, $
 the symmetry of the  modular matrix $S_{\zl \zm}\,$ is recovered.

Combined with \Ij{b}, equations \Vg, \Vh, \Vi\ imply that the matrix
elements $(N_a)_b^c\,,$ $b,c \in T\,,$ of the subset
of $N$  matrices, $\{N_a\,, a\in T\}\, $ coincide with the
extended Verlinde multiplicities  and hence are nonnegative
integers.  One can show that
 $(N_a)_b^c =0 $ for $a,b \in T\,, c \not \in T\,,$
 because  all $N$ matrix elements
are assumed nonnegative and
 $k_a\,$ are  strictly positive. (Hint: Use also that $k_a\,, a\in
\CV\,,$ and $ k_c\,, \, c\in T\,$
provide linear representations of
the graph algebra and the extended Verlinde algebra respectively.)

 We are thus in the position to use the results of sect. 5.1.
The
matrices $\{N_a\,, a\in T\}\, $ span  a C - subalgebra $\cU_T$ of
$\cU$, isomorphic to the extended Verlinde algebra. The
latter  is selfdual, hence
 the subalgebra $\cU_T$ coincides with its dual algebra -- a quotient
of the dual algebra  $\hat{\cal U}$.
Indeed the  subset $\hat T$ is identified with the
set of weights in the
class of identity $\{\zr \}\,$ and $\hat{\cU}_{\hat T}\,$ is
realised as a subalgebra  of the $M$
matrices. The splitting of the  set of exponents
 into the equivalence classes $\{\zl\}$,
each described by  the decomposition of
a representation of the extended algebra,  agrees   with the
meaning of the equivalence relation in the sense of \BI.

Let us rewrite the  basic relation \sm, using \Ve\ and \Vg, as
\eqn\Vk{
N_{\{\za\} \{\zb\}}^{\{\zg\}} \ \sqrt{k_c\over k_{a} \ k_{b}}
=\sum_{\zg \in \{\zg\}}\ M_{\za \zb }^{\zg} \,
\sqrt{\hat{k}_{\zg}\over \hat{k}_{\za} \  \hat{k}_{\zb}}
\,, \quad a= \{\za\}\,, b= \{\zb\}\,, c= \{\zg \} \in T \,,}
identifying the equivalence classes $\hat{T}_i\, $ of
$\hat{\cU}\,$ with
the representations $\{\zl\}\,$ of the extended algebra.
Then \Vk\   can be  seen as
 an expression  for  the structure constants
 $N_{\{\za\} \{\zb\}}^{\{\zg\}}$ of the  quotient algebra
 $\hat{\cU}/\hat{\cU}_{\hat T}\,$
 in terms of the structure constants $M$
of $\,\hat{\cU}\,.$ The parameter $\hat{k}_{\{\zg\}}
  =\sum_{\zg\in \{\zg\}}\,
\hat{k}_{\zg}/ \sum_{\za\in \{\un\}}\,
 \hat{k}_{\za}  = D^2_{\{\zg\}}\, $
coincides with $k_c\,$ for $c\in T\,, c\equiv \{\zg\}\,.$
%
%
%
\medskip\penalty-500\nind
{\bf Remarks}\penalty10000
\item{1)} Note that the relations of the type in \Vg,
\Vh, follow from the general results of \BI;  specific to
the present application is the symmetry of the unitary matrix in the
r.h.s. of \Vg, \Vh, and furthermore, the explicit realisation in \Vdb.
\item{2)} Similarly there is a  formula dual to \Vk, in which
 the $M$ - matrix elements are replaced by  $N$ -- matrix elements,
 the sum runs over $c\in T_c\,,$ etc, and
 the l.h.s. is replaced by the
structure constants of $\cU/\cU_T\,,$
dual to the subalgebra
$\hat{\cU}_{\hat T}\,,$ where ${\hat T}$ coincides with the
identity class $\{\zr\}\,.$
It leads to the dual counterparts of the relations \Vg, \Vh.
However the  analog of the extended $S$ matrix, which diagonalises
the  matrices representing the algebra $\cU/\cU_T\,,$
(as well as the matrices  $(M_{\za})_{ \zb}^{\zg}\,,$ $ \, \za\,,
\zb\,, \zg \in \{\un\}$)
is not necessarily  a symmetric matrix in
general. Typically one has  furthermore  to enlarge the subset ${\hat T}\,,$
i.e., to consider subalgebras of the $M$ - algebras, bigger than
the identity subalgebra $\hat{\cU}_{\{\zr\}}\,,$ so that the
automorphism \Iaa\
keeps ${\hat T}\,$ invariant.  The structure constants of
the factor algebra $\cU/\cU_T\,,$ are not in general integer valued.
\item{3)}
 The relevance of the $C$--algebras  for the study of the
relation between  graphs and  nondiagonal modular
invariants   was first pointed out in \DFZ.  The considerations
above are to some extent inverse in spirit to what was done
in \DFZ\ for the block-diagonal cases of $sl(2)$ and $sl(3)$.
 Starting from the explicitly constructed graphs and $N$-algebras
for these examples, it was there  observed empirically
that in all cases
the  extended fusion algebra is represented isomorphically by a
subalgebra of the graph algebra $N$.  This leads to the relation
\Vh.  It was furthermore argued,  exploiting the
approach of Bannai and Ito, that
given a subalgebra of a type I graph algebra, described by a
subset $T$ of the vertices, there is a natural dual partition of
the set of exponents, which yields a type I modular invariant.
In the present approach the knowledge of the modular invariant
is assumed, from which it is {\it derived} that the graph algebra admits a
subalgebra isomorphic to the extended Verlinde one. Furthermore
the relation \sm, which we traced to originate from the locality
requirements in the field theory, provides an additional
information and yields in particular  the explicit general
expression \Vdb\ for the subset $\{\psi_a^{(\zl)}\}\,, a \in
T\,,$ in terms of the  original
$S_{\zl \zm }\,$ and the extended $S_{\{\zl \} \{\zm \} }\,$
modular matrices.
\item{4)}
For completeness let us indicate what is the explicit set $T$
for each of the graphs in the figures of the present paper.
For  $sl(3)$, (see also \DFZ),  the graphs of Fig 1 and 4 below
have been drawn in such a way that the vertices of $T$ lie
at the periphery of the graph: the respectively 4, 6 and 3 external
vertices form the set $T$. For the graph of Fig. 2, they
are the vertices labelled 1 to 6, and for that of Fig. 3, those
labelled 1 to 10.

%
%
%
\subsec{ Other  restrictions on the  eigenvectors $\psi_a\,$ }
\nind
According to property ix) in section 2  and to
the discussion following eq. \IIja\ the fundamental
adjacency matrix $G_1$ always coincides with  one of the $N$ matrices,
$(G_1)_b^c= N_{a_f b}^c\,.$
This enlarges the
set of $|T|$ explicit $1$ - dim linear representations of the $M$
- algebra in \Vv, adding a  new  one. Indeed we have
${\psi_{a_f}^{(\zl)} \over \psi_{\un}^{(\zl)} }= \zg_1^{(\zl)}=
{S_{\GLh_1+\zr \, \zl} \over  S_{\zr\, \zl}}\,,$ cf. \Ii, and
since the denominator $\psi_{\un}^{(\zl)} $  is known from \Vdba,
this determines  $\psi_{a_f}^{(\zl)}\,$ and  gives for the
corresponding  eigenvalues of $M_{\zl}$ the explicit expression
\eqn\fe{
{\psi_{a_f}^{(\zl)} \over \psi_{a_f}^{(\zr)}}
={\zg_1^{(\zl)}\over \zg_1^{(\zr)}}\, \sqrt{D_{\{\zl\}} \,D_{\zl}}
= {S_{\GLh_1+\zr \ \zl} \over  S_{\GLh_1+\zr\  \zr}}\,
\sqrt{D_{\{\zl\}} \over D_{\zl}}
\,.}
%
Now given \fe\ we add to the system \sm\ the equations
\eqn\esm{
{\psi_{a_f}^{(\zl_1)} \over \psi_{a_f}^{(\zr)} }\,
{\psi_{a_f}^{(\zl_2)} \over \psi_{a_f}^{(\zr)} }=
\sum_{\zl_3 \in \Exp}\, M_{\zl_1\, \zl_2}^{\zl_3}\,
{\psi_{a_f}^{(\zl_3)} \over \psi_{a_f}^{(\zr)} }\,.}
 Similarly the complex conjugated eigenvector
$\psi_{a_f}^{(\zl_1) *}\,$ is expressed through the eigenvalues
of the matrix $G_{n-1}\,.$
As an illustration of the application of \fe, \esm, consider the
example $\slh{(3)}_9 \subset \widehat{E_6}_{_{1}}$ with modular invariant
\eqn\invc{
Z_{\slh{(3)}_9}=
|\chi_{_{1,1}}+\chi_{_{1, 10}}+\chi_{_{10,
1}}+\chi_{_{5,5}}+\chi_{_{2,5}}+\chi_{_{5,2}} |^2+ 2
|\chi_{_{3,3}}+
\chi_{_{6,3}}+\chi_{_{3,6}}|^2\,.}
Accounting for \sym, \aut, the system \sm\ reduces in this case
to $3$ independent equations for $6$ unknown
matrix elements $M_{(5,5) (5,5)}^{(5,5)}\,, \, M_{(5,5)
(5,5)}^{(2,5)}\,, \, M_{(3,3)_+ (3,3)_+}^{(3,3)_-}\,,$
 $\, M_{(3,3)_+ (3,3)_+}^{(3,6)_-}\,,$ $ \, M_{(3,3)_+
(3,3)_-}^{(5,5)}\,,$ $   M_{(3,3)_+ (3,3)_-}^{(2,5)}\,.$ (Here
$\big((3,6)_{\pm}\big)^* = (6,3)_{\mp}\,,$ etc.; $\pm $
correspond to the representations $(2,1,1,1,1,1)\,, (1,1,1,1,2,1)\,$ of
$\widehat{E_6}_{_{1}}\,.$   Since the
extended fusion algebra is of type $\IZ_3$, i.e.,  any product
contains only one representation, the summation in \esm\ runs
effectively always over $\zl_3\,$ within one class,
$\{\zl_3\}\,.$ Then \esm\  adds $3$ new equations for the $6$ variables
and thus allows to determine completely all $M$ matrices,
and to reconstruct the graph denoted
$\CE_1^{(12)}$ in \DFZ\ (Fig. 4).
%
%
\fig{The graph  $\CE^{(12)}_1$ corresponding to the conformal embedding
$\slh(3)_9\subset (\hat{e}_6)_1$}
{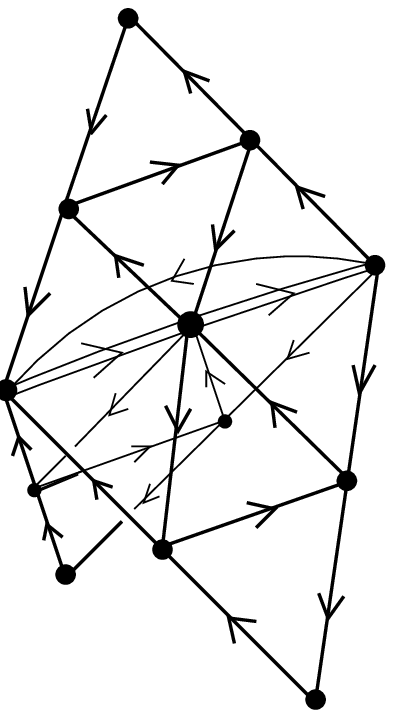}
{3cm}\figlabel\trifol

 In exactly the same way one can solve also the case
 $\slh(4)_6 \subset \slh(10)_1\,$ (see  \SYa\
  for the expression of
the modular invariant), thus exhausting, together with the example
discussed in Appendix A, the cases   $\slh(4)_k \subset \ggh\,,$
all described in \IIIa.

 More information on the eigenvectors can be obtained if we
 make the assumption that
for  any vertex $a$ belonging to the subset $T\,$, the graph of $G_1$ has
only one edge starting from $a$ and one edge ending at $a\,$.  This extra
condition seems to be satisfied by graphs associated with conformal
embeddings, although we do not know a proof of it. This implies  that
$G_1\, N_a =N_{b(a)}\,$ for any  $a \in T$ and some vertex
$b(a)$  and hence
\eqn\Vl{
{\psi_{b(a)}^{(\zl)}\over \psi_{\un}^{(\zl)}}=
{\psi_{a_f}^{(\zl)}\over
\psi_{\un}^{(\zl)}} \, {\psi_{a}^{(\zl)}\over \psi_{\un}^{(\zl)}}
\,, \qquad \forall a \in T \,.}
It is easy to see
that all these vertices $b(a)$, $a\in T$ form the class $T_{a_f}\,.$
Using  again \Vd, \Ve, we obtain new explicit eigenvalues of the
$M$ matrices and hence new equations for their matrix elements.

 The set of explicitly found $\psi_b^{(\zl)}$ is furthermore
enlarged by the complex conjugates of the above vectors,
since they generate the class $T_{\av_f}\,.$ ($T_{a_f}\,$
and $T_{\av_f}\,$ may coincide.)

 Examining the graphs on Figs. 1,2,3 we see that
 \Vdb, \fe\ and  \Vl\ (and their complex conjugates) together
provide  explicit analytic
expressions for all eigenvectors $\phi_a\,, $ $a \in \CV\,.$
These formulae are not sufficient for the graph on Fig.4
associated with the invariant \invc\  for which one recovers the
eigenvectors corresponding to all but the
three ``central" vertices. These vertices have different values
of the $\IZ_3$ grading $\tau(a)\,$  introduced in section 2. This allows
to express the  moduli of the
corresponding eigenvectors  (and in particular
the remaining unknown elements of the Perron--Frobenius eigenvector
$\psi^{(\zr)}$) in terms of the known ones with the
same grading, by  exploiting the following relations:
\eqn\gd{
\eqalign{&\sum_{ b, \tau(b)=l} |\psi_b^{(\lambda)}|^2 ={1\over
n}\,, \quad l=0,1,
..., n-1 \,, \cr
&\forall \zl\in \Exp\,, \ \  {\rm such\ \  that } \quad
\gamma_p^{(\lambda)} \not =0\,,
p=0,1,..., n-1\,.}}
This follows from the basic property  \Ig\ of the matrices $G_p\,$
since
\eqn\gdo{
\zg_{_{p}}^{(\zl)} \,\sum_{ b, \tau(b)=l} \psi_b^{* (\nu)}\,
\psi_b^{(\lambda)} =\sum_{a, b; \tau(b)=l}\, \psi_b^{* (\nu)}\,
(G_p)_{b a}\, \psi_a^{(\lambda)}=
\zg_{_{p}}^{(\nu)} \,\sum_{ a, \tau(a)=l-p} \psi_a^{* (\nu)}\,
\psi_a^{(\lambda)}\,}
 for any $p=1,2,...,n-1\,.$ Choosing
$\zn=\zl$ and taking into account the first relation in \Ik{a}\
we get in particular \gd.

\bigskip

As the reader might have already noticed, the graphs drawn in
Figs. 1--4 (all corresponding to conformal embeddings) possess
some symmetry. Let us try to trace  the origin of these symmetries.

Some of the symmetry  comes from the  $\IZ_2$ group generated by
the charge conjugation of the vertices, i.e., it reflects \Iga\ and the
first equality in \Ik{b}.

Further restrictions on the set  $\psi_a^{(\zl)}\,,$  and hence on
the $N$ -- matrices, result from the   group $\zG_{\ggh}\,$ of
automorphisms (to be denoted $\zS\,$)  of the Dynkin diagram of
 $\ggh$, generated by automorphisms 
which do not fix the vertex corresponding to the affine root.

The induced action of $\zG_{\ggh}\,$ on the integrable weights of $\ggh$
can be lifted to the set of vertices $\CV$. Indeed the subset $T$,
being in one-to-one correspondence with the set of
integrable  weights of the extended algebra  $\ggh$, remains invariant
under the action of these  automorphisms.
Using \Vdb\  we have for $ a\in T\,$
\eqn\Ba{
\psi_{\zS(a)}^{(\zl)} =e^{2 i \pi Q_{\zS}(\{\zl\}) }\,
\psi_a^{(\zl)} \,,
 \quad \zl\in \{\zl\}\,,}
where  the phase $Q_{\zS}(\{\zl\})$ is specific  to the
 extended algebra $\ggh$  and satisfies $Q_{\zS}(\{\zr\})=0\,:$  see e.g.
\GRW\ for details.   Moreover \Vl\  allows
to extend this action to the elements of the class $T_{a_f}\,, $
i.e., to define $\psi_{\zS(b(a))}^{(\zl)}:=\zg_1^{(\zl)}\,
\psi_{\zS(a)}^{(\zl)}= \psi_{b(\zS(a))}^{(\zl)}\,, $ for any
$a\in T$. In fact \Ba\ extends to an arbitrary vertex $a$ as well,
taking into account that  $M_{\zl \zm }^{\zg} $  can be nonzero
only if $Q_{\zS}(\{\zg\})=Q_{\zS}(\{\zl\})+Q_{\zS}(\{\zm\})\,$
mod $\IZ\,,$  since the same is true for the extended multiplicity
$N_{\{\zl\} \{\zm\}}^{\{\zg\} }\,.$
Furthermore it is not difficult to show
that the invariance of $T$ under $\zS$ implies the invariance of
any of the classes $T_i\,,$ since due to the above definition we
have  $N_{a \zS(b)}^{\zS(c)}= N_{a b}^c\,,$
and  this is true in particular for $b\in T\,, $ $a \sim c\,.$
 The equality \Ba\ implies that for the fixed points of $\zS\,$
the elements $\psi_a^{(\zl)}\,$ vanish identically whenever
$Q_{\zS}(\{\zl\})\,$  is nontrivial.

 Combining \Ba\ and \quar\ we have for any $a$ and $\zl$ such that
$\psi_a^{\zl}\not =0\,, $
\eqn\cons{
{l\over n}\, \Big(\tau(\zS(a))- \tau(a)\Big)=
 Q_{\zS}\big(\{\zs^l(\zl)\}\big) - Q_{\zS}(\{\zl\})\,, \quad {\rm
mod}\ \ \IZ\,.}
The consistency of this  relation  restricts in general the automorphisms
$\zS$ to some subgroup of $\zG_{\ggh}\,.$

\medskip

Examples:

1.  $k+3=12$ for $sl(3)$. The extended algebra is
$\hat{E_6}$ and   the group $\zG_{\ggh} \,$ is isomorphic to
$\IZ_3$. The latter is realised by the
fusion of the set of integrable representations of $\hat{E_6}$
for $k=1$. One has
 $Q_{\zS}(2,1,1,1,1,1)=-1/3=- Q_{\zS}(1,1,1,1,2,1)$ (see, e.g.,
\GRW\ ). The classes
 $T\,, T_{a_f}\,, T_{a_f^{\vee}}\,,$ consist each of three elements,
while the remaining three classes
consist of one element, denoted, say,   $a=4,8,12\,,$
respectively -- on Fig.4 these are the three interior vertices.
 This implies
in particular that if $\zl \in \{(3,3)_{\pm}\}\,,$ then
$\psi_a^{\zl}=0$ for  $a=4,8,12\,, $
since $ \psi_{\Sigma^l(a)}^{(3,3)_{\pm}}= e^{\mp 2\pi i l \over 3 }\,
\psi_a^{(3,3)_{\pm}} =\psi_a^{(3,3)_{\pm}}$.
%
\medskip
2. The case $\slh(5)_3 \subset \slh(10)_1$: here \Ba\ and \cons\ make
sense for $\zS$ in the subgroup $\IZ_5 \subset \IZ_{10}$,
consisting of the even powers of the generating element $\zS^1\,$ defined
as in  \Ia\ for $n=10$; $Q_{\zS^{2s}}(\{\zl\})= s\, t(\{\zl\})/5\,,$
where $t(\{\zl\})\,$ is the standard $n$ - ality of the weight $\{\zl\}\,.$
  In general the symmetry \Ba\ implies
the invariance $(G_p)_{\zS(a) \zS(b)}=(G_p)_{a b}$.
 To check it in the case  of Fig. 3 requires  in particular the
identification of vertices in $T$ with the integrable weights of $\slh(10)_1$.
The analysis of sect. 5.2 applied to this case reveals
that the first six fundamental representations of $sl(10)$
correspond to vertices labelled
 1, 8, 5, 7, 4 and 6 in Fig \rosebud;
their conjugate are consistent with the involution $a\mapsto a^\vee$.
It follows that $\Sigma^2\,$ maps cyclically vertices
  $1\to 5 \to 4\to 3 \to 2\to 1$ and $ 6\to 10\to 9\to 8
\to 7 \to 6$
as well as $ 11 \to 15\to 14 \to 13 \to 12 \to 11$ and $16 \to 20 \to 19\to
18 \to 17\to 16 \,.$
Accordingly the graphs on Fig.3 are invariant under rotations by $2\pi/5\,.$

%
%


%
\subsec{The intertwiner }
\nind
 We point out in this subsection a simple  application of the
general formula \Vdb.

 In \DFZ\ an intertwiner, which relates the adjacency
matrices of the given graph  and of the basic ``$A$-''graph
associated with $sl(n)$ ($n=2,3$)  with the same value of $h$,
 was constructed, namely
\eqn\Vm{\eqalign{
&N_{\GLh_p+\zr}\, V= V\,G_p\,,\cr
 &V_{\zg b}^c = \sum_{\zo \in \Exp}\,
{S_{\zg \zo}\,\psi^{(\zo)}_b\, \psi^{(\zo)\, *}_c
\over S_{\zr \zo} }\,,  \qquad b,c \in \CV\,, \ \ \zg\in
 {\CP}^{(h)}_{++} \,. \cr }}

It was furthermore  {\it observed} that in all type I examples
of $n=2,3$, the
particular matrix elements $ V_{\zg 1}^c\,, \ c \in T\,,$ encode
the content of
the blocks $\{\zg\}\,.$
 This property of the intertwiner \Vm\ may
be now {\it derived} as a consequence of the explicit expression
\Vdb, namely
\foot{ This has also been derived recently
 by Ocneanu \Ocn\ in a different context as
reflecting the counting of ``essential paths'' on the graph.}
\eqn\Vn{  V_{\zg 1}^c =\sum_{\zO}\,
S_{c  \zO}^*
\sum_{\zo \in \zO}\,
S_{\zg \zo} \,
={\rm mult}_{\zG}(\zg)\,
  \zd_{
\zG c }\,, \qquad c \in T\,,}
i.e.,   $V_{\zg 1}^c$ provides the multiplicity of $\zg$ in $
\zG \equiv c$.
 Here
$S_{\zg \zo} $ is the ordinary $S$ matrix whereas
 $ S_{c  \zO}\equiv S_{\zG \zO} \, $ refers to the extended one;
we deviate from  our usual notation, introducing
$\zO$ instead of $\{\zo\}$ for a representation of the
extended algebra: this is to avoid confusion in cases where
some $\zo$ belongs to several $\zO$.
The second equality in \Vn\ is the standard consistency condition
resulting from the modular properties of the character  $\chi_{_{\zO}}=
\sum_{\zo \in \zO}\,   \chi_{_{\zo}}=
\sum_{\zo \in
 {\CP}^{(h)}_{++}}\, {\rm mult}_{\zO}(\zo) \,  \chi_{_{\zo}}\, .$

More generally, if $b,c\in T\,,$ we have
\eqn\Vp{ V_{\zg b}^c = \sum_{\zG \ni \zg} {\rm
mult}_{\zG}(\zg)\,
N_{\zG b}^{c }
 \,,}
where $N_{\zG b}^{c }\,$ is the extended Verlinde multiplicity,
and hence, at least  for $b,c\in T$, $V_{\zg b}^c$  are nonnegative integers
(in particular zero for $\zg \not \in \Exp$). To get \Vp\  insert for $\psi_a
^{(\zl)}\,$ the solution  \Vdb\  and
use again the consistency condition from the modular
transformation of the characters in the form
\eqn\Vr{\sum_{\zo}\, {\rm mult}_{\zO}(\zo)\,
S_{\zg \zo} =\sum_{\zG \ni \zg }{\rm mult}_{\zG}(\zg)\,  S_{\zG \zO} \,,}
or, alternatively, insert \Vn\ in $V_{\zg b}^c=\sum_a \,V_{\zg 1}^a\,
N_{a b}^c\,$; for $b,c \in T$ this sum restricts to $T$.
In the cases with   no degeneracy of weights $\zo$,
coming from different representations of the extended algebra, the sum
in \Vp\ reduces to one term and hence
$ V_{\zg b}^c  = {\rm mult}_{\{\zg\}}(\zg)\, N_{\{\zg \} b}^{c}\,.$

%
%
%
%

\newsec{ Conclusions.}
\nind
In this paper we have presented more evidence
on the connection between conformal theories and graphs.
This connection and a number of related facts that had so far remained
empirical have received more support or have been
proved to follow from natural assumptions.

Building upon our previous work on $sl(2)$ theories in
which we had established some relations between data on the OPA
and data relative to the graphs,  we have extended
these relations to higher rank $sl(n)$ theories.
Our present work is restricted to conformal theories  that we call
of type I, for which the block diagonal form of the modular
invariant signals the existence of an extended chiral algebra.
Two plausible assumptions on the consistency of the operator
algebra with this
feature, namely eqs \fa\ (in the cases of conformal embeddings)
and \class, have led us to the important
relation \sm\ between fusion multiplicities, quantum dimensions and
structure constants of the Pasquier algebra constructed out of the
graphical data.

The validity of these assumptions and the
practical importance of this relation \sm\ have been tested in  a variety of
cases of CFT associated with conformal embeddings.
 It has been shown that
they allow in some cases to fully determine the graphs starting
from the CFT. Thus new graphs have been obtained in a way that is
much more systematical than the empirical procedures used so far.
The relation \sm\ for the Pasquier algebra is presumably more
general, as suggested
by the  example   in App. B, and applies to all type I cases.

We have also reconsidered the use of the theory of $C$-algebras in
connection with our problem. Merging this approach with
the previous results provides a justification or a new perspective
to results that had been obtained some time ago in \DFZ. Moreover
an explicit general formula \Vdb\ has been  derived
for the components
of the  eigenvectors $\psi_a\,,$ corresponding to the special
set of vertices $T\ni a\,,$
that are in one-to-one correspondence with the representations of the
 given   extended theory.

 What remains to do is to understand better the justification
of the assumptions of sect. 3. Any progress in the solution of
the general duality equations would eventually allow to extend
  the system of equations \sm\  and determine completely  the Pasquier
algebra, as was done in  \PZ\ in the $sl(2)$ case.
Also cases with  nontrivial multiplicities within a given
representation of the extended algebra seem to present new
situations that have just been tackled in App. A. Furthermore 
the extension of these results to theories of type II, i.e. that are
not block diagonal, presents a  challenge.
In that respect the recent work of Ocneanu \Ocn\
seems to indicate that these theories may also be treated in a
similar way and that there are interesting connections between the graphs
pertaining to the pair of theories of type I and II obtained from one
another by a twist.

\medskip

Finally we notice that the new graphs found in this work yield
new cases of infinite reflection groups, following the procedure
of \Zub, and describe presumably
 patterns of solitons in $\CN=2$ supersymmetric theories, as
discussed by Cecotti and Vafa
\ref\CV{S. Cecotti and C. Vafa, {\it Comm.Math.Phys.} {\bf 158} (1993) 569.}.

  \bigskip

{\bf Acknowledgements}
\medskip
We would like to thank M. Bauer,  A. Coste and P. Di Francesco
for stimulating conversations.
V.B.P. acknowledges the warm hospitality  of the theory group at
Saclay and  the financial
support of Commissariat \`a l'Energie Atomique, the hospitality
of ASI, TU Clausthal,  and a partial support of the Bulgarian
Foundation for Fundamental Research under contract $F - 404 -
94\,.$

\bigskip

%
%
%
%

\appendix{A}{An example of nontrivial multiplicity }
\nind
In this Appendix we shall illustrate on the example $\slh(4)_4
\subset \soh(15)_1\,$ the cases
when an exponent $\zl$ appears with a multiplicity higher than $1$
in a given representation of the extended algebra.

The modular invariant associated with this embedding reads \SY\ :
\eqn\Aa{
Z_{\widehat{sl}(4)_4}=
|\chi_{_{111}}+\chi_{_{151}}+\chi_{_{123}}+\chi_{_{321}}|^2 +
|\chi_{_{115}}+\chi_{_{511}}+\chi_{_{212}}+\chi_{_{232}}|^2 +
|2\chi_{_{222}}|^2\,.}
The exponents in the three blocks correspond to the content in
the decomposition of the three integrable representations
$(1,1,1,1,1,1,1)\,, (2,1,1,1,1,1,1)\,, $ and $(1,1,1,1,1,1,2)\,$
 of  $\soh(15)_1\,$
with quantum dimensions respectively $1\,, 1\,,$ and $\sqrt{2}\,.$
They close  on an Ising type fusion algebra. The representation
$(2,2,2)$ of $\slh(4)_4$ appears in
$(1,1,1,1,1,1,2)\,$  with a multiplicity $2$.  Thus unlike the
previous examples the
extended algebra does not
distinguish the  two fields to be associated with the exponent
$(2,2,2)$.
On the other hand if we identify these fields, i.e., look at them
as two copies of one and the
same field,  the set of equations for the $M$ -- matrices is checked
to be inconsistent. To resolve the ambiguity of the representations
$(2,2,2)$, which is a fixed point under the standard action of
 automorphism group $Z_4$,  introduce two fields
$ \Big((2,2,2);\pm \Big)$ and define charge conjugation and the
action of the $\sigma$ automorphism according to
\eqn\resol{
\sigma\Big((2,2,2);\pm \Big)=\Big((2,2,2);\mp \Big)\,, \quad
\Big((2,2,2);\pm \Big)^* = \Big((2,2,2);\mp \Big)\,,}
i.e., each of the two fields is invariant under $\sigma^2$ only.
For the other exponents keep the standard definitions of the charge
conjugation and  of $\sigma$.
Then  the last term in the modular invariant \Aa\ can be interpreted as
$|\chi_{_{(222);+}}+\chi_{_{(222);-}}|^2 \,.$

The definition of the charge conjugation implies that
$M_{\((2,2,2);\pm \)\((2,2,2);\mp \)}^{(1,1,1)}=1$.
With \resol\ the set of  equations for the $M$ -- matrices  becomes
consistent and we obtain
using that $D_{(2,2,2)}= 4(1+\sqrt{2})\,, \quad D_{(1,2,3)} =
(1+\sqrt{2})^2\,,$
\eqn\Ab{
M_{(1,2,3) (3,2,1)}^{(1,2,3)}={1\over 2}\,
\Big(\sqrt{D_{(1,2,3)}}-
{1\over \sqrt{D_{(1,2,3)}}}\Big) =1\,, }
\eqn\Ac{
M_{\big((2,2,2);\pm \big)\big((2,2,2);\pm
\big)}^{(1,2,3)}={1\over 2 D_{(1,1,1,1,1,1,2)}} \, {D_{(2,2,2)}
\over \sqrt{D_{(1,2,3)}} }= \sqrt{2}\,,}
\eqn\Ad{
M_{\big((2,2,2);\pm
\big)\big((2,2,2);\mp\big)}^{(1,2,3)}=\Big({1\over 2
D_{(1,1,1,1,1,1,2)}} \, {D_{(2,2,2)} \over
\sqrt{D_{(1,2,3)}} } - {1\over \sqrt{D_{(1,2,3)}}  } \Big)=1\,.}
 The remaining matrix elements are either zero
or determined from  \sym, \aut, taking into account \resol.

\medskip
Diagonalising the $M$ matrices we obtain the fundamental matrices $G_p\,$
\def\sq{\sqrt{2}}
 described by graphs depicted in Fig. 5
%
%
%
\fig{The graphs of $G_1$ and $G_2$ corresponding to the conformal embedding
$\slh(4)_4\subset \soh(15)_1$. The graphs should be viewed as
having a $\IZ_2\times \IZ_2$ symmetry : $1\leftrightarrow 2$ and
$4\leftrightarrow 5$. The broken lines denote edges carrying
$\sq\,$.  The set $T$ is made of vertices 1,2 and 3.}
{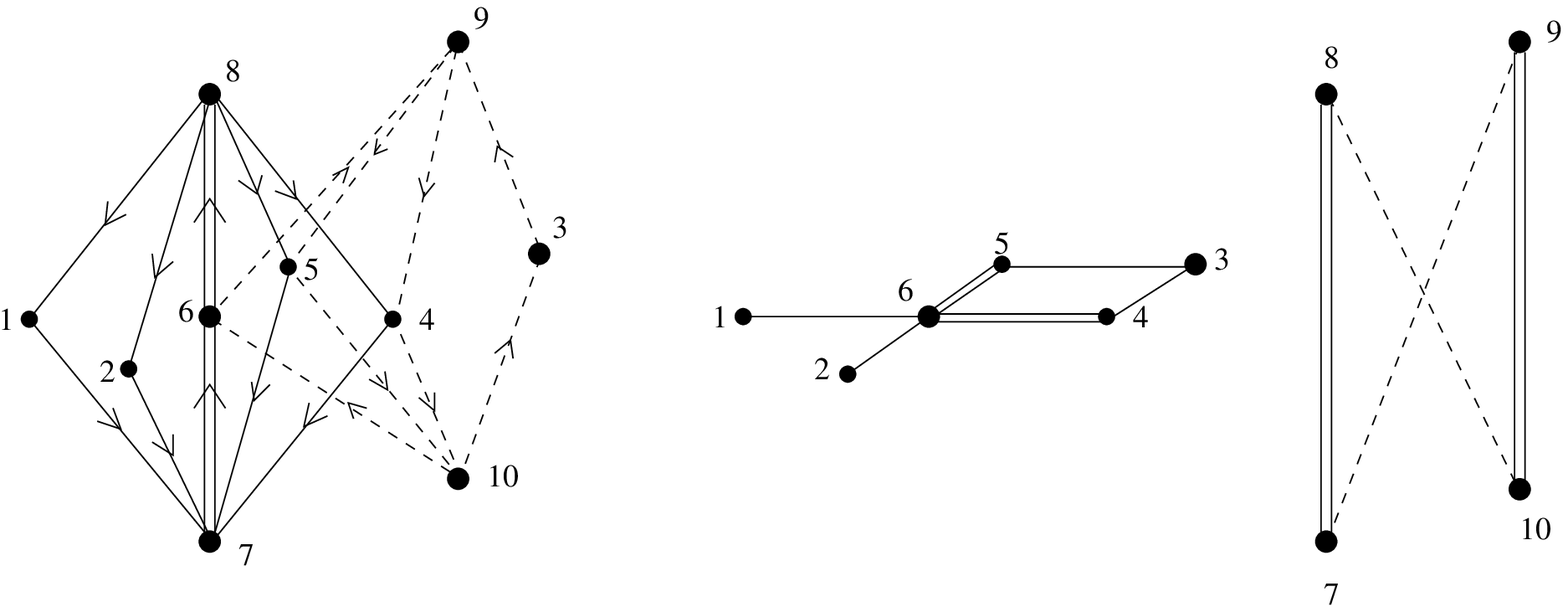}{12cm}\figlabel\chall

An unexpected
feature is that these matrices have non integral entries~!!
This is reminiscent of what is obtained when one folds a
$\IZ_2$ symmetric graph: for example going from the $A_{2n-1}$
Dynkin diagram to its $\IZ_2$ quotient $B_n$, one finds that
the (symmetrized) Cartan matrix of the latter contains some
$\sq$  entries. This suggests that the graphs above may be unfolded
into graphs with 10+2=12 vertices.

In fact if one returns to the modular invariant \Aa\ and calls
``exponents" the weights that label the diagonal terms of the
sesquilinear form in the characters, one finds {\it twelve}
such exponents. In that standpoint  the exponent $(2,2,2)$ should come
with a multiplicity 4.

%
%
%
\fig{The same graphs as in \chall\ after `unfolding', i.e.
duplication of vertices 9 and 10 into 9, $9'$, and 10, $10'$.}
{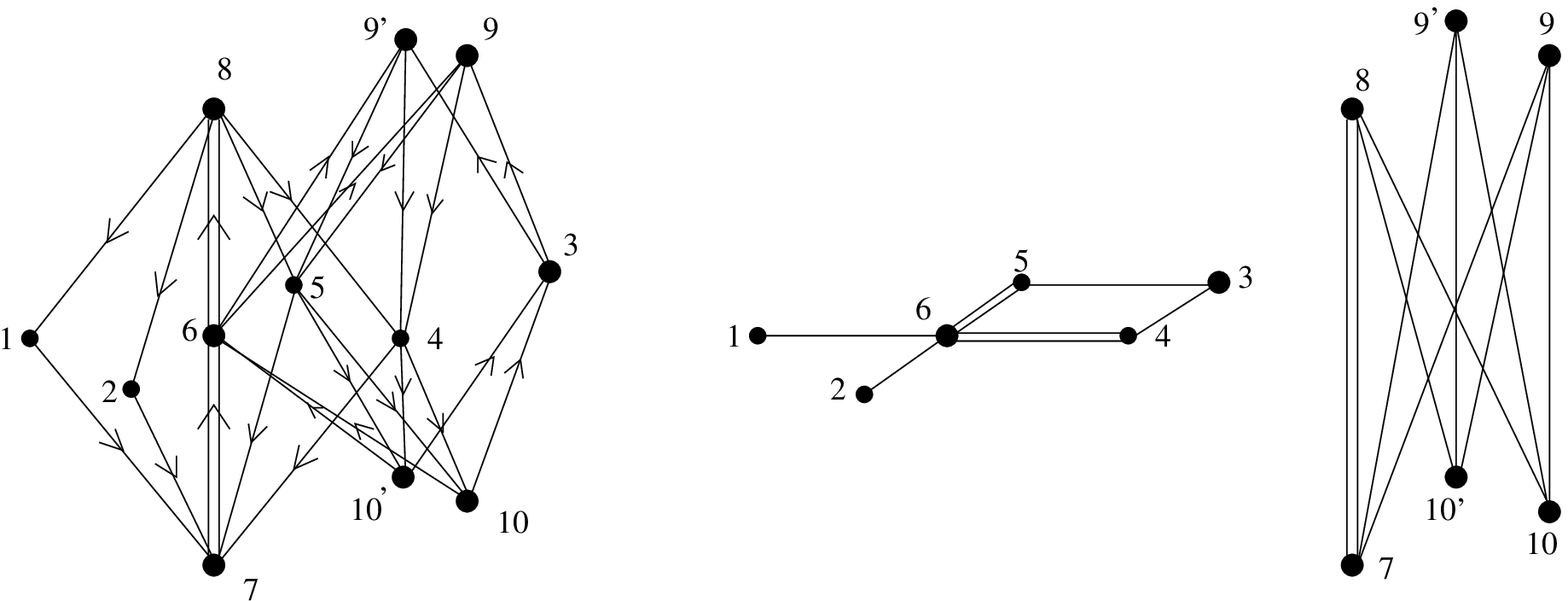}
{12cm}\figlabel\defi

There is indeed a simple way of unfolding the graphs above.
One just duplicates the vertices denoted 9 and 10 into $9,9'$
and $10,10'$ and
one replaces in $G_1$ and $G_2$
the entries $\sq$ by $(1,1)$ or $\pmatrix{1\cr 1\cr}$
and the entry $2$ by a $2\times 2$ block $\pmatrix{1&1\cr 1&1\cr}\,,$
 see the resulting graphs on Fig. 6. The
exponent $(2,2,2)$ now appears
 4 times.
 It seems, however, that it is impossible to make all the
structure constants of the $M$ and $N$ algebras non negative.

It thus appears that in such a  case with
 nontrivial  multiplicities, there are two alternative attitudes:
\item{--} either one puts the emphasis on the extended algebra,
i.e. the block structure of $Z$, with all the consequences
that may be inferred on the $\psi$'s, as discussed in the present paper,
but at the expense of dealing with non integrally laced graphs, 
not covered by the considerations of sect. 2;
\item{--} or one follows the usual scheme of attaching an
exponent to each diagonal term of $Z$, and constructs an
integrally laced graph, but at the expense of a more
complicated connection with the OPA.

\appendix{B}{More on orbifolds of $sl(3)$ }

\nind This appendix is devoted to another series of CFT's
to which our considerations  apply: the so-called $\CD$ series of
orbifold $\slh(3)$ theories for levels $k=0\ \mod 3$.
In contrast with the case mentioned above in \Ikor, they
are endowed with a block modular invariant partition function \B\
\eqn\Baa{Z={1\over 3}\sum_{\Gl\in \CP^{(k+3)}_{++} \cap Q}
| \chi_{\Gl}+\chi_{\Gs(\Gl)}+\chi_{\Gs^2(\Gl)}|^2\,, }

where we recall that $Q$ denotes the root lattice. The partition function
\Baa\ is as usually normalized in such a way that $|\chi_\Gr|^2$ comes with a
factor 1.  The term $|\chi_ {\Ga_0}|^2$ associated with the fixed
point $\Ga_0=({k+3\over 3}) \Gr$ of
$\Gs$ \Iaa\ comes however with a
multiplicity 3. Accordingly, we have to append an index $i=1,2,3$ when
referring to the exponent $\Ga_0$.

For the $\slh(3)$ orbifolds under consideration, the relevant
graphs have been constructed long ago by Kostov \Ko. They are
obtained by an orbifolding procedure from the basic graphs of type $\CA$
and the simplest example has been displayed in Fig. 1. They have
a number of vertices
equal to
$$ {k\over 2} ({k\over 3}+1) +3\ . $$
In the same way as in the case of the $D_{{\rm even}}$ orbifolds of $\slh(2)$
theories in which two subseries have to be distinguished, depending
whether ${k\over 2}=0 $ or $ 2\ \mod 4$, here we have to distinguish according
to the value of ${k\over 3}\ \mod 3$.
This manifests itself in  particular in the form of
the extended $S$ matrix. That matrix  satisfies \Vr, but
this leaves some arbitrariness in the $3 \times 3 $ block
$S_{\{(\za_0\,, i) \} \{(\za_0\,, j) \}}\,.$   The remaining elements are
determined from \Vr, using that all exponents
have $\tau=0$, and hence exploiting \quar\  gives
 $S_{\{\zl \} \{\zm \}}=3 S_{\zl \zm}$
for $\zl, \zm \not = \za_0\,,$ $S_{\{\zl \} \{(\za_0\,, i )\}}= S_{\zl \za_0}$
for $\zl \not = \za_0\,.$
From the Kac -- Peterson formula for $S_{\zm \zn}\,$ it follows that
$S_{\za_0 \za_0} = -3/h$ for $h=k+3=6$ mod $9\,,$
$S_{\za_0 \za_0} = 0$ for $h=0$ mod $9\,,$ $S_{\za_0 \za_0} =
3/h$ for  $h=3$ mod $9\,.$  Also note $S_{\za_0 \zr}=3/h\,.$
The sum of the entries on any line or
row in the $3 \times 3$ block in the extended matrix should be equal to
$S_{\za_0 \za_0} $ according to \Vr, i.e., accounting also for the
symmetry we are left with 3 unknown parameters, to be determined
e.g., by the unitarity condition, or the other conditions on the
extended modular matrix. It appears that if we call
$\kappa$
the integral part of $(k-3)/9$, this $3\times 3$ block
has the following form
\eqn\Bc{\Big( S_{\{(\Ga_0,i)\} \{(\Ga_0,j)\}}\Big)=
{3\over h}\pmatrix{{k\over 3}-\Gk&-\Gk-1&-\Gk-1\cr
-\Gk-1 &{k\over 3}-\Gk & -\Gk-1\cr
-\Gk-1& -\Gk-1 &{k\over 3}-\Gk \cr}\ . }
 The charge conjugation of the classes $ \{(\Ga_0,i)\}\,$ is
assumed trivial,
$\{(\Ga_0,i)\}^* = \{(\Ga_0,i)\}\,,$  $i=1,2,3$ for all
three subseries.

From the relations between the S matrices of the original and of
the extended theories, we deduce that
$D_{\{(\za_0,i)\}} ={D_{\za_0}\over 3}$
and $D_{\{\zl\}} \equiv D_\zl$ otherwise. Then \sm\ reads:
\eqn\Ca{
\sum_{l=0}^2\, M_{\zl \zm}^{\zs^l(\zg)} = N_{\{\zl \} \{\zm
\}}^{\{\zg\}}\,, }
for $\zl, \zm, \zg$ all different from the fixed point $\za_0\,;$
\eqn\Cc{
\sqrt{3}\, M_{\zl \zm}^{(\za_0\,, i )} = N_{\{\zl \} \{\zm
\}}^{\{\za_0\,, i \}}\,, }
for $\zl, \zm\,$  different from the fixed point;
\eqn\Cb{
{1\over 3} \,\sum_{l=0}^2\, M_{(\za_0\,, i) (\za_0\,,j)}^{\zs^l(\zg)}
=  M_{(\za_0\,, i) (\za_0\,,j)}^{\zg} = N_{\{(\za_0\,, i) \} \{(\za_0\,, j)
\}}^{\{\zg\}}\,, }
for $ \zg$  different from the fixed point (here we have used
also \aut\ );
\eqn\Cd{
  M_{(\za_0\,, i) (\za_0\,, j)}^{(\za_0\,, l)} = \sqrt{3} \,
N_{\{(\za_0\,, i)
\} \{(\za_0\,, j) \}}^{\{(\za_0\,, l)\}}\,. }
Given the extended modular matrix
one can   compute  by Verlinde formula the
extended multiplicities and insert in these formulae.

All these relations as well as \Vdb\ have indeed been checked for
the lowest representatives of the three subseries, $k=3,6,9$ and
it is thus presumed that the equations \sm\ and \Vdb\ hold true
in general for all these orbifold theories.   Furthermore the
equations \sm\ and \Vdb\
are expected  to hold true  for the general $\slh(n)\,$ orbifolds of \Ko.

\bigskip

\listrefs

\bye